\documentclass[
prl,
twocolumn,
showpacs,
superscriptaddress,
preprintnumbers,
bibnotes,
amsmath,
amssymb,
aps
]{revtex4-1}

\usepackage{amsmath}
\usepackage{enumitem}
\usepackage{graphicx}
\usepackage{dcolumn}
\usepackage{bm}
\usepackage[flushleft]{threeparttable}
\usepackage{amsmath}
\usepackage{amssymb}
\usepackage{braket}
\usepackage{multirow}
\usepackage{array}
\usepackage{url}
\usepackage{booktabs}
\usepackage{float}
\usepackage{hyperref}
\usepackage{mathtools}
\usepackage{setspace}
\usepackage{lipsum}
\hypersetup{colorlinks=true, urlcolor=blue, citecolor=cyan, pdfborder={0 0 0},}
\newcolumntype{P}[1]{>{\centering\arraybackslash}p{#1}}

\usepackage{color,soul}
\newcommand{\yk}[1]{{\color{black}{{#1}}}} 
\newcommand{\nn}{\nonumber \\} 
\newcommand{\vv}[1]{{\boldsymbol #1}} 
\newcommand{\bpm}{\begin{pmatrix}}
\newcommand{\epm}{\end{pmatrix}}
\newcommand{\fig}[1]{Fig.\,\ref{#1}}
\newcommand{\figs}[1]{Figs.\,\ref{#1}}
\newcommand{\figr}[1]{Figure\,\ref{#1}}

\makeatletter
\newcommand{\Rom}[1]{\expandafter\@slowromancap\romannumeral #1@}
\makeatother

\begin{document}
\title{Dual topological nodal line and nonsymmorphic Dirac semimetal in three dimensions}    

\author{Yun-Tak Oh}
\affiliation{Department of Physics, Sungkyunkwan University, Suwon 16419, Korea}
\author{Hong-Guk Min}
\affiliation{Department of Physics, Sungkyunkwan University, Suwon 16419, Korea}
\author{Youngkuk Kim}
\email{youngkuk@skku.edu}
\affiliation{Department of Physics, Sungkyunkwan University, Suwon 16419, Korea}
\date{\today}       

\begin{abstract}
Previously known three-dimensional Dirac semimetals (DSs)  occur in two types -- {\em topological} DSs and {\em nonsymmorphic} DSs. Here we  present a novel three-dimensional DS that exhibits both features of the {\em topological} and {\em nonsymmorphic} DSs.  
We introduce a minimal tight-binding model for the space group 100 that describes a layered crystal made of two-dimensional planes in the $p4g$ wallpaper group. 
Using this model, we demonstrate that double glide-mirrors allow a noncentrosymmetric three-dimensional DS that hosts both symmetry-enforced Dirac points at time-reversal invariant momenta and twofold-degenerate Weyl nodal lines on a glide-mirror-invariant plane in momentum space. 
The proposed DS  allows for rich topological physics manifested in both topological surface states and topological phase diagrams, which we discuss in detail. 
We also perform first-principles calculations to predict that the proposed DS is realized in a set of existing materials BaLa$X$B$Y_5$, where $X$ = Cu or Au, and $Y$ = O, S, or Se.
\end{abstract}
\maketitle

Dirac semimetals (DS) refer to a class of topological semimetals, characterized by hosting massless Dirac fermions in momentum space \cite{RevModPhys.90.015001}. 
First identified in graphene with the vanishingly weak spin-orbit coupling (SOC), the massless Dirac fermion system has attracted a surge of interest, exhibiting exotic properties and potential applications for future electronic devices \cite{geim2009graphene,allen2009honeycomb}. 
Notably,  with the advent of topological insulators \cite{Hasan10p3045, qi2011topological}, the three-dimensional (3D) DS with strong spin-orbit coupling has reinforced their status as an important class of topological semimetals.
It was first noted that a 3D DS can occur at the phase boundary between the topological and the normal insulators in the presence of inversion symmetry \cite{07fu-kane-mele, murakami-07-njp}.
Later, Young \textit {et al.} found that the 3D DS can be stabilized by crystalline symmetries and time-reversal symmetry \cite{12young}, and Wang {\em et al.} theoretically proposed the material realizations in Na$_3$Bi and Cd$_3$As$_2$ \cite{12wang, 13wang}, which were confirmed experimentally \cite{14Liu_science, 13xu_arxiv,14borisenko_prl, 14neupane_nat.com.,14Liu_nat.mat.}.
Currently, the DSs are expected to exist in a variety of forms, such as two-dimensional (2D) DSs \cite{16wieder}, double DSs \cite{16-wieder-prl}, type-{\Rom 2} DSs \cite{Chang2017prl}, and Dirac-Weyl semimetals \cite{PhysRevLett.121.106404}.

In spite of this variety, it is surprising to notice that all the previously known DSs fall into two disjoint classes, dubbed {\em topological} and {\em nonsymmorphic} DSs, respectively \cite{14yang}.
The nonsymmorphic class of the DSs is characterized by hosting Dirac points (DPs) that are pinned at the time-reversal invariant momenta (TRIMs) of the Brillouin zone (BZ). 
On the other hand, the topological class of the DSs distinguish themselves from the nonsymmorphic DSs by having a pair of DPs off TRIMs. 
Another distinguishing feature of the topological DSs is the coexistence of nontrivial band topology in the bulk, manifested as gapless  excitations on the surface \cite{kargarian16pnas,bednik18prb}. In contrast, the bulk bands of the nonsymmorphic DSs are expected to be topologically trivial.  Instead, a topological nature of the nonsymmorphic class is reflected in topological phase transitions, driven by symmetry-lowering perturbations from the nonsymmorphic DS into either a topological insulator or a normal insulator \cite{12young,14steinberg,14yang,schoop16natcomm,17yang_prb}. 

In this paper we provide an exception to this {\em a priori} classification of 3D DSs. 
Developing a minimal tight-binding model for space groups (SGs)  $P4bm$ (\# 100), we establish the existence of  a novel type of 3D DSs, characterized by featuring both the {\em topological nodal lines} and {\em nonsymmorphic} DSs. It is shown that the DS hosts the DPs that reside at TRIMs, which is a characteristic feature of the nonsymmorphic DSs. Simultaneously, the bulk bands carry nontrivial band topology, giving rise to topological surface states, which is unexpected from the previously known nonsymmorphic Dirac semimetals.
A striking consequence of this dual {\em nonsymmorphic} and {\em topological} nature of the DS is the rich topological physics manifested not only in the surface energy spectrum but also in topological phase transitions driven by symmetry-breaking perturbations.
Drumhead-like topological surface states arise due to the nontrivial band topology in the bulk, characterized by hosting Weyl nodal lines (WNLs). 
Moreover, symmetry-lowering perturbations derive a topological phase transition from the proposed DS to distinct topological phases, including a weak topological insulator (WTI) and Weyl and double Weyl semimetal (WS) phases. Using first-principles calculations, we also discuss its material realization in  an existing compound, BaLaCuBO$_5$.


\begin{figure}[tb]
\includegraphics[width=0.48\textwidth]{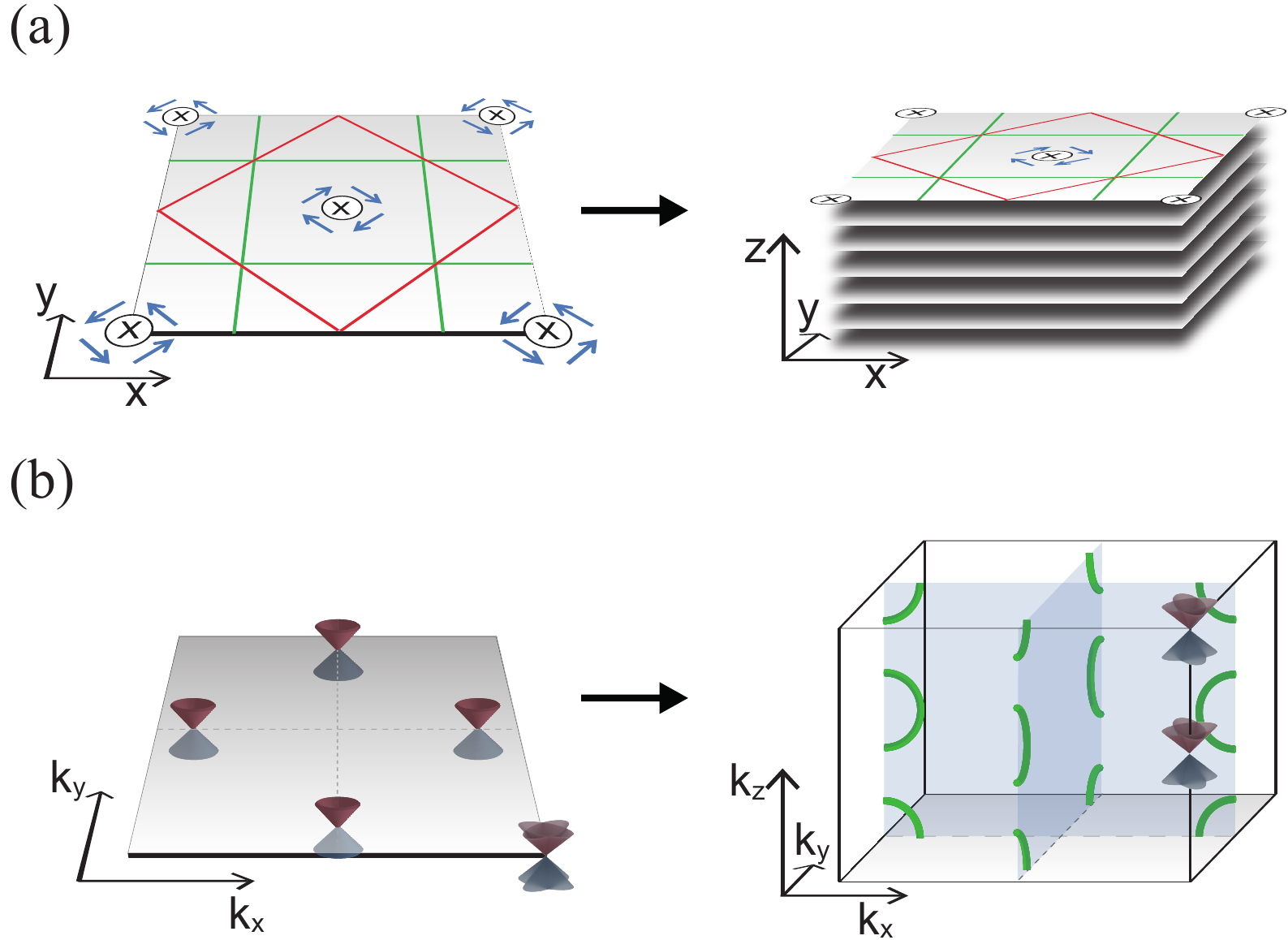}
\caption{\label{fig:Stacking}  
(a) Schematic illustration of a two-dimensional (2D) layer in the $p4g$ wallpaper group (left panel) and an infinite stack of the $p4g$ layers (right panel). The red (green) lines represent mirror (glide-mirror) invariant lines. The center of $C_{4z}$ rotational axis is designated by the $\otimes$ symbol. (b) Corresponding BZ in two (left) and three (right) dimensions. The location of the twofold-degenerate WPs and fourfold-degenerate DPs are indicated by the Weyl and the double Weyl (Dirac) cones, respectively. The location of WNLs in the 3D BZ is indicated by green lines.
}
\end{figure}

Let us begin with elucidating the role of symmetries in SG 100 to protect degeneracies of the Bloch states. SG 100 has the distinguishing feature that it is generated by a glide-mirror $g_x$ and a fourfold rotation $C_{4z}$ without inversion symmetry. As emphasized in \cite{zaheer14thesis, wieder-18-science}, the double glide-mirrors, $g_x$ and $g_y =  C_{4z}^{-1} g_x C_{4z}$, together with time-reversal symmetry $\mathcal{T}$, span four-dimensional irreducible representations (FDIRs) at $M = (\pi,\pi, 0)$ and $A=(\pi,\pi,\pi)$, where $g_x$ and $g_y$ satisfy the minimal algebras for a FDIR, $g_x^2 = g_y^2 = 1$ and $[\mathcal{T},g_{x(y)}]  = \{g_x,g_y\} = 0$.  Moreover, the linear dispersion of the bands is generic at $M$ and $A$ since a $\mathcal{T}$-odd vector representation of the point group at the $M$ and $A$ points is present in the tensor product of the FDIRs \cite{12young, zaheer14thesis, 16-wieder-prl}. Therefore, the presence of DPs are enforced in SG 100 when the filling is an odd multiple of four.  In addition, $g_x$ ($g_y$) and $\mathcal{T}$ further give rise to a  constraint to the connectivity of the bands, such that the Kramers pairs at $\Gamma$ and $Y(X)$ should exchange their partners from $\Gamma$ to $Y(X)$ without opening a band gap, leading to hourglass-like connectivity \cite{young15prl, wang16nature,wieder-18-science}. As a consequence, it is guaranteed that additional twofold-degenerate WNLs are present on the $k_x = 0$ ($k_y = 0$) plane, protected by glide-mirror $g_x$ ($g_y$). 

\begin{figure}[tb]
\includegraphics[width=0.48\textwidth]{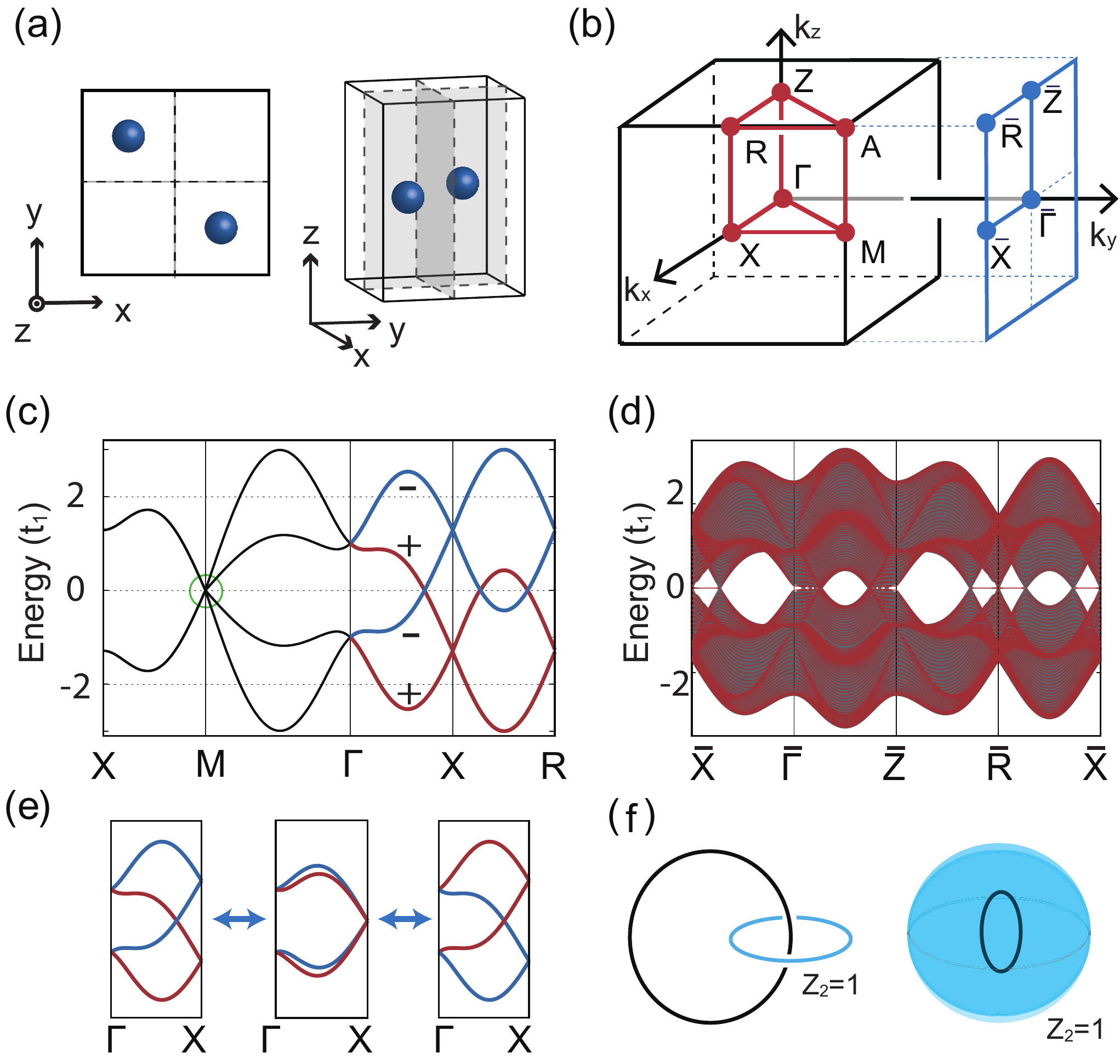}
\caption{\label{fig:tb-4b} (a) Model lattice for SG 100. 
The glide planes are represented by dashed boxes.
(b) Bulk tetragonal and surface rectangular BZs.
(c) Electronic energy bands for SG 100, calculated from (\ref{eq:tb-full}) with the parameter set $\{t_1,t_2, t_3,v_0, v_1, v_2, v_3, v_4\}  = \{0.35, 0, 0, 0, 0.5, 0.6, 0.45\}$.
The bands in the $+$ and $-$ eigensectors of $g_y$ ($g_x$) are colored by red and blue, respectively.  
The DP is indicated by a dashed (green) circle. 
(d) Bulk (grey) and slab (red) energy bands. Topological surface states emerge in the interior region of the projected nodal lines, where $\overline{X}$ and $\overline{R}$ are contained. 
(e)  Schematic illustration of band inversion at $X$ ($R$). (f) Topological characterization of twofold-degenerate nodal lines in SG 100.
}
\end{figure}

The above symmetry-analysis provides a guiding principle to design the DS hosted in SG 100. 
Since SG 100 and the $p4g$ wallpaper group are equivalent, generated by $C_{4z}$ and $g_x$, a minimal four-band tight-binding model can be constructed from an infinite stack of the identical layer in the $p4g$ wallpaper group as illustrated in \fig{fig:Stacking}. 
The constructed lattice model is presented in \fig{fig:tb-4b}(a). 
A unit cell comprises two sublattices $A$ and $B$ (labeled by $\tau_z = \pm1$), which are coordinated at  $\vv d(\tau_z) = \frac{1}{4} \left[\left(2 + \tau_z \right){\vv a}_x + \left(2-\tau_z \right){\vv a}_y\right]$, respectively.
The corresponding tight-binding Hamiltonian is given as 
\begin{equation}
\mathcal{H}^0( \vv k )  = \mathcal{H}^t (\vv k)  + V (\vv k), 
\label{eq:tb-full}
\end{equation} 
where
\begin{align}
\mathcal{H}^t({\vv k}) = \, &  t_1 \cos \frac{k_x}{2} \cos\frac{k_y}{2} \tau_x \nn
 + & t_2 (\cos k_x + \cos k_y) + t_3 \cos k_z
\nonumber
\end{align}
describes the nearest hopping of electrons, and
\begin{align}
V(\vv k) =  \, & v_0 \cos \frac{k_x}{2} \cos \frac{k_y}{2} \tau_y \sigma_z + v_1 \left( \sin k_x \,  \sigma_x + \sin k_y\,  \sigma_y \right) \tau_z \nn
+ &  v_2\left(\sin k_x \, \sigma_y - \sin k_y \, \sigma_x \right) +  v_3 \sin{k_z} \tau_z \sigma_z \nn 
 + & v_4 \left( \sin  \frac{k_x}{2} \, \cos \frac{k_y}{2} \sigma_y 
- \cos  \frac{k_x}{2} \sin  \frac{k_y}{2} \sigma_x \right) \tau_x  \nonumber
\end{align}
describes the potential terms that lower the transnational symmetry of $\mathcal{H}^t({\vv k})$ into SG 100. 
$V(\vv k)$ is constructed, such that it preserves the generators of SG 100, $ g_x = \tau_x \exp\left( -i \frac{\pi}{2} \sigma_x \right)$ and $C_{4z}  =  \exp\left( -i \frac{\pi}{4} \sigma_z \right)$, and time-reversal symmetry $\mathcal{T} = i \sigma_y K$, where $\{\sigma_i\}_{i=x,y,z}$ are the Pauli matrices for spins.
We adopted a gauge, in which the Hamiltonian $\mathcal{H}^0(\vv k)$ transforms under the translation of a reciprocal lattice vector $\vv G$ according to
\begin{equation}
\mathcal{H}^0(\vv k + \vv G)   =  e^{-i \vv d(\tau_z) \cdot \vv G} \mathcal{H}^0(\vv k) e^{i  \vv G \cdot \vv d(\tau_z)}. \nonumber 
\end{equation}

\begin{figure*}
\centering
\includegraphics[width=0.95\textwidth]{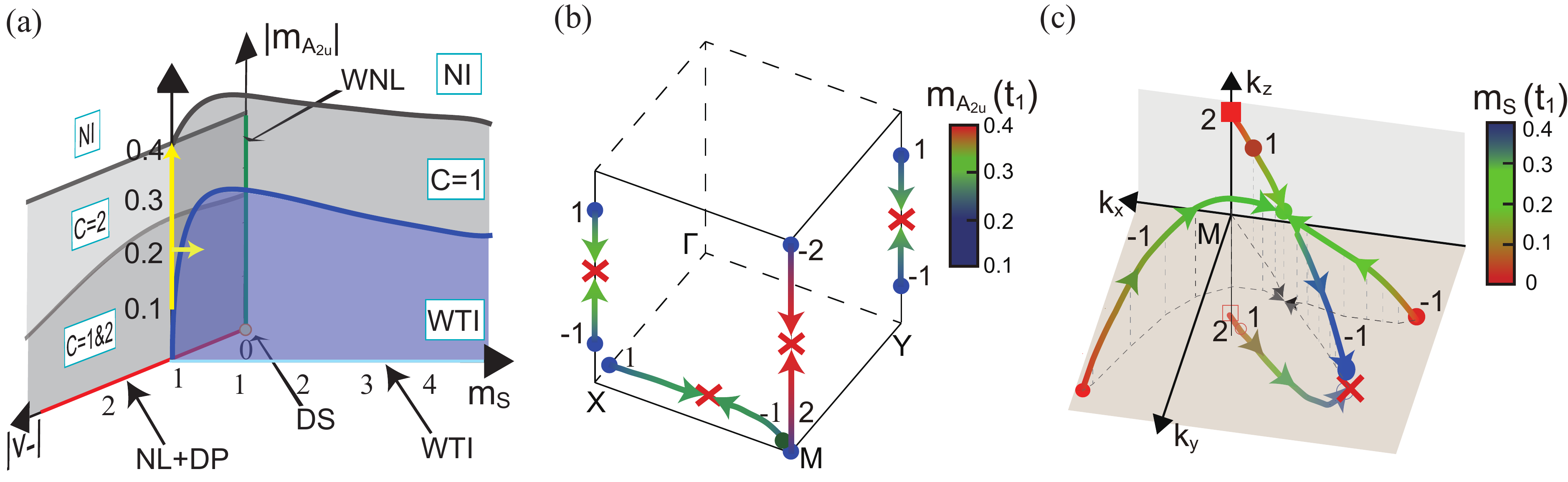}
\caption{
Topological phase diagram and topological phase transitions induced by symmetry-lowering perturbations from the DS in SG 100. 
(a) Topological phase diagram in the ($v_{-}, m_s, m_{\protect\scalebox{.55}{$A_{\protect\scalebox{1.}{$2u$}}$}}$)-parameter space in the unit of $t_1$.   
The DS in SG 100 resides along the $|v_-|$ axis (colored by red), which is connected with a centrosymmetric DS at the origin, where  $(v_{-},  m_s, m_{\protect\scalebox{.55}{$A_{\protect\scalebox{1.}{$2u$}}$}}) = (0,0,0)$ (indicated by a yellow circle). 
A WNL semimetal appears along the $|m_{\protect\scalebox{.55}{$A_{\protect\scalebox{1.}{$2u$}}$}}|$ axis (green line).
The WS phases are present 
in the grey-colored regions on the $m_s-|m_{\protect\scalebox{.55}{$A_{\protect\scalebox{1.}{$2u$}}$}}|$ and $|v_{-}|-|m_{\protect\scalebox{.55}{$A_{\protect\scalebox{1.}{$2u$}}$}}|$ planes. 
The WS phase carrying both double ($|\mathcal{C}|=  2$) and single ($|\mathcal{C}| = 1$) WPs is distinguished from the WS phase carrying only double ($|\mathcal{C}|=  2$) WPs with thicker gray color on the $|v_-|-|m_{\protect\scalebox{.55}{$A_{\protect\scalebox{1.}{$2u$}}$}}|$ plane.
A WTI phase is colored by blue on the $m_s-|m_{\protect\scalebox{.55}{$A_{\protect\scalebox{1.}{$2u$}}$}}|$ plane.
The remaining white area represents the trivial insulator phase.
(b) Evolution of the WPs as a function of $m_{\protect\scalebox{.55}{$A_{\protect\scalebox{1.}{$2u$}}$}}$ during the topological phase transition occurring along the vertical yellow line in (a).
The numbers near a trajectory indicate the corresponding Chern number of the WPs.  
A color scheme is used to indicate the magnitude of $m_{\protect\scalebox{.55}{$A_{\protect\scalebox{1.}{$2u$}}$}}$ at $m_\mathrm{s}$ = 0 and $v_- = 1.0$. 
The WPs are annihilated at the red crosses. 
(c) Evolution of the WPs as a function of $m_s$ during the topological phase transition occurring along the horizontal yellow-line in (a).
A square (circle) represents a $|\mathcal{C}| = 2$ ($|\mathcal{C}| = 1$) WP.
} \label{fig:phase_diagram}
\end{figure*}

\figr{fig:tb-4b}(c) shows the electronic energy bands calculated from the tight-binding model Eq.\,\ref{eq:tb-full}. 
Without the inversion symmetry, each band along the high-symmetry $M$-$\Gamma$-$X$-$R$ line is non-degenerate, thus forming a fourfold degeneracy at $M$. 
Note that the bands are linearly dispersing in the vicinity of $M$ point. 
Therefore, the bands feature a nonsymmorphic DP at $M$. We also have confirmed that an additional DP is present at $A$, as we expected from the symmetry-analysis.
Based on the Wilson bands calculations \cite{11yu}, we find that the DPs carry the zero Chern number \footnote{See the Supplemental Materials for the details of the Wilson bands calculations.}.
This indicates that the fourfold degeneracy is a genuine DP, in the sense that it is a composite of two WPs with $\pm 1$ Chern numbers, respectively.  

Besides the fourfold degeneracy, we find that the bands also feature twofold-degeneracy WNLs, the hourglass-like band connectivity on the high-symmetry $\Gamma$-$X$\,($Z$-$R$) line. This  guarantees the presence of a twofold degeneracy on the $k_y = 0$ plane, as shown in \fig{fig:tb-4b}(c). 
A close inspection in the entire BZ reveals that one-dimensional nodal lines are present in the vicinity of $X$\,($R$) lying on the $k_y = 0$ plane. We have confirmed that a Weyl line node carries the $\pi$ Berry phase, calculated along a $C_2\mathcal{T}$-invariant path that threads the nodal line [See the left panel of \fig{fig:tb-4b}(f).], where the Berry phase is $\mathbb{Z}_2$-quantized. As a consequence of the $\pi$ Berry phase, drumhead-like states emerge on the surface where the projected interior region of a nodal line has non-zero area, such as the (100) surface. As shown in \fig{fig:tb-4b}(d), the slab band calculation results in the topological surface states at $E = 0$ near the  $\overline{\Gamma}$\,($\overline{R}$), which constitute a part of the drumhead-like surface states on the (010) surface.  

The WNL hosted in SG 100 is of a hourglass-type  \cite{bzduek16nature, l.wang17prb, wang17nat.comm.}, which is robust against the band inversion at $M$\,($R$). As illustrated in \fig{fig:tb-4b}(e), the band inversion at $X$ shrinks the size of the nodal line into a fourfold-degenerate DP. However, instead of annihilating it, the band inversion reverts the DP to a nodal line due to the hourglass-like band connectivity. We assert that this type of WNLs can be  characterized by a non-trivial $\mathcal{Z}_2$ topological invariant, calculated on the time-reversal invariant sphere that encloses a nodal line [See the right panel of \fig{fig:tb-4b}(f).]. The Wilson bands calculation results in the same connectivity of the Wilson bands as those of a 3D Dirac point \footnote{See the Supplemental Materials for the detailed results of the topological invariants.}. The nontrivial $\mathcal{Z}_2$ invariant, again, reveals that the WNL can be shrunk to form a 3D DP. 

Having demonstrated the {\em topological} aspect of the DS, we now move to its {\em nonsymmorphic} aspect, captured in a topological phase diagram shown in \fig{fig:phase_diagram}.
From the DS phase, we consider symmetry-lowering perturbations \footnote{See the Supplemental Materials for the classification of the perturbations by the point group D$_{\rm 4h}$.}.
Among diverse possibilities, as a representative example, here we consider a combination of the inversion symmetric ${E_g}$- and $B_{2g}$-mode strains and an $m_{\scalebox{.55}{$A_{\scalebox{1.}{$2u$}}$}}$-mode staggered potential.
These perturbations are described by a perturbed Hamiltonian $\mathcal{H}^1 (\vv k)$, where
\begin{align}
\mathcal{H}^1 (\vv k) = & m_{\scalebox{.55}{$E_{\scalebox{1.}{$g$}}$}} \sin\left( \frac{k_x+k_y}{2}\right) \tau_y  + m_{\scalebox{.55}{$B_{\scalebox{1.}{$2g$}}$}} \sin\frac{k_x}{2}\sin\frac{k_y}{2} \tau_x  \nn
 & + m_{\scalebox{.55}{$A_{\scalebox{1.}{$2u$}}$}} \tau_z. \label{eq:h-pert-part}
\end{align}
For simplicity, we assume the mass parameters are equivalent between the inversion-symmetric perturbations ($m_s \equiv m_{\scalebox{.55}{$E_{\scalebox{1.}{$g$}}$}} = m_{\scalebox{.55}{$B_{\scalebox{1.}{$2g$}}$}}$).
Furthermore, we decompose the pristine Hamiltonian $\mathcal{H}^0(\vv k)$ (Eq.\,\ref{eq:tb-full}) into the inversion-symmetric part $\mathcal{H}^{0}_+(\vv k)$ and inversion-asymmetric  part $\mathcal{H}^{0}_-(\vv k)$, where
\begin{align}
\mathcal{H}^{0}_+=& t_1 \cos\frac{k_x}{2} \cos\frac{k_y}{2} \tau_x + v_1 \left(\sin k_x \tau_z \sigma_x + \sin k_y \tau_z \sigma_y \right)  \nn
& + v_3 \sin{k_z} \tau_z \sigma_z,
\end{align}
and 
\begin{align}
\mathcal{H}^{0}_- =\, &  v_- \Bigg[ v_0 \cos\frac{k_x}{2} \cos\frac{k_y}{2} \tau_y \sigma_z +  v_2\left(\sin k_x \, \sigma_y - \sin k_y \, \sigma_x \right)   \nn
 +& v_4 \left( \sin  \frac{k_x}{2} \, \cos \frac{k_y}{2}  \sigma_y  - \cos  \frac{k_x}{2} \sin  \frac{k_y}{2}  \sigma_x \right)\tau_x  \Bigg].
\end{align}
Here, $v_-$ is introduced to parametrize the overall strength of the inversion-asymmetric part.

\figr{fig:phase_diagram}(a) shows a topological phase diagram that is obtained from $\mathcal{H}^0 + \mathcal{H}^1$  in the ($v_-$,$m_s$,$m_{\scalebox{.55}{$A_{\scalebox{1.}{$2u$}}$}}$) space. We first note that the DS phase in SG 100 resides along the $|v_-|$ (red) axis.
From this DS phase, a centrosymmetric strain, described by $m_s$, drives a topological phase transition;  positive (negative) $m_s$ induces a WTI (normal insulator), characterized by $\mathbb{Z}_2$ topological indices  $(\mu_0;\mu_1,\mu_2,\mu_3) = (0;001)$[(0;000)]. Therefore, the DS phase defines a phase boundary between the normal and topological insulator phases \yk{\cite{murakami-07-njp}}, thus exhibiting the nonsymmorphic nature of the DS.
In addition to the WTI phase, we find that a Weyl semimetal (WS) can also be induced from the DS  phase by applying the staggered potential ($|m_{\scalebox{.55}{$A_{\scalebox{1.}{$2u$}}$}}|>0$).
Interestingly, we find that the three distinctive WS phases are allowed: (1) one having regular (single) WPs with the Chern number $|\mathcal{C}| = 1$, (2) another having double WPs with $|\mathcal{C}| = 2$, and (3) the other having both single and double WPs. 
We also note that an archetypal centrosymmetric DS phase is restored from the DS phase by turning off the noncentrosymmetric interactions $v_- = 0$, from which a WNL semimetal phase is induced by the $m_s$ strains, represented by a vertical green line in the figure. 

Figure\,\ref{fig:phase_diagram}(b) illustrates the detailed process of topological phase transition via the creation and annihilation of WPs along the vertical (yellow) path indicated in Fig\,\ref{fig:phase_diagram}(a).  
When varying $m_{\scalebox{.55}{$A_{\scalebox{1.}{$2u$}}$}}$ from 0.1 to 0.4 in the unit of $t_1$, the in-plane $\mathcal{C}=1$ WP near $X$ ($Y$) and the in-plane $\mathcal{C}=-1$ WP near the DP of $M$ fuse and annihilate eventually,
while the WPs residing on the $k_z$-axis find their anti-chiral partners by moving along the $k_z$ axis.
This inter-TRIM WP annihilation results in the trivial insulator phase.
On the other hands, \fig{fig:phase_diagram}(c) illustrates the evolution of the WPs during the topological phase transition from the WS with $\mathcal{C} = 2$ to the WTI phase that occurs along the horizontal (yellow) path indicated in Fig.\,\ref{fig:phase_diagram}(a).
Apart from zero, $m_s > 0$ splits a WP with $\mathcal{C}=2$ into two $\mathcal{C}=1$ WPs off the $k_z$-axis. One of the two $\mathcal{C}=1$  WPs encounters with other two $\mathcal{C}=-1$ WPs from the $k_z = 0$ plane. 
This event results in a single $\mathcal{C}=-1$ WP, indicated by a solid green circle. 
The resultant $\mathcal{C}=-1$ WP is eventually annihilated on the $k_z = 0$ plane by meeting with another WP with $\mathcal{C}$=1, which originates from the double WP on the $k_z$-axis. This annihilation results in a WTI \footnote{See Supplemental Material at http:// for the detailed calculations of the associated topological invariants.}.


\begin{figure}[tb!]
\includegraphics[width=0.48\textwidth]{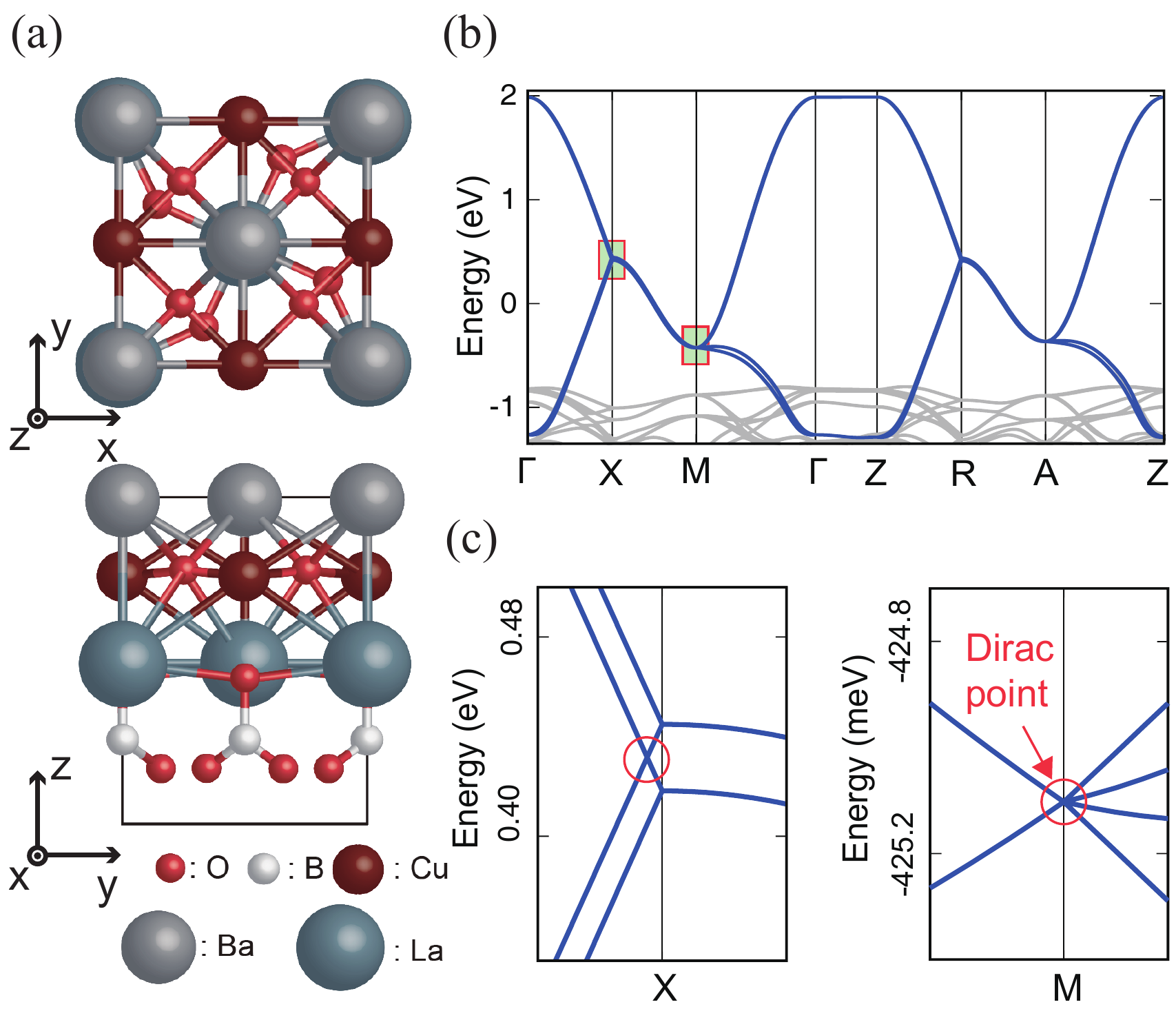}
\caption{\label{fig:dft} Atomic and electronic structures of BaLaCuBO$_5$. 
(a) Top view (top panel) and side view (bottom panel) of the atomic structure.  (b) Electronic energy band structure.  The unitcell is indicated by a solid box. 
The DFT (Wannier) energy bands are colored by grey (blue).  The Wannierzation exactly reproduce the four DFT bands near the Fermi level. (c) Magnified views of the red rectangles in (b) at $X$ (left panel) and at $M$ (right panel).  The WNL and DP are indicated by a red circle.
}
\end{figure}

Finally, searching for materials that realize the DS in SG 100, we have found an existing material BaLaCuBO$_5$ \cite{Norrestam1994}.
BaLaCuBO$_5$ is a layered system in SG 100 as shown in \fig{fig:dft}(a). It comprises $p4g$ multilayers with an each layer preserving  $C_{4z}$ rotation and double glide-mirrors $g_x$ and $g_y$ symmetries. Our first-principles calculations\yk{, performed using \textsc{Quantum Espresso} package \cite{giannozzi09p395502},} support that BaLaCuBO$_5$ realizes the proposed DS phase \footnote{See  Supplemental Material at http://xxxxx for the details of computational methods and  the first-principles results for other candidates.}. \figs{fig:dft}(b) and \ref{fig:dft}(c) show the first-principles electronic energy bands of BaLaCuBO$_5$.
The sticking of four bands is clear from the band structure\yk{, featuring filling-enforced gaplessness \cite{watanabe2016}}. A fourfold-degenerate DP is present at $M$, and the hourglass-like band connectivity appear on the $\Gamma$-$X$ line. The hourglass-like band connectivity leads to a band crossing on the $\Gamma$-$X$ line, as shown in the magnified view in \fig{fig:dft}(d). The presence of band crossing signals the presence of a WNL that encircles the $M$ point lying on the $k_x = 0$ plane, which we have confirmed throughout the band calculations performed in the entire BZ. Our results is in good agreement with the a time-reversal
invariant topological encyclopedia online, which indicates BaLaCuBO$_5$ as a high-symmetry point topological semimetal \cite{Zhang19p475}.

In conclusion, we have established the existence of a novel type DSs in three dimensions, characterized by hosting topological surface states and mediating topological phase transitions.  Hosting topological surface states in the nonsymmorphic DSs, the proposed DS features a unique topological character unlike archetypal 3D DSs. The surface energy spectrum should  give rise to  drumhead-like topological surface states, which should be feasible to observe in the BaLaCuBO$_5$ compound using a known experimental technique, such as angle-resolved photoemission spectroscopy (ARPES). Moreover, defining a symmetry-tuned topological critical point between a normal insulator and a WTI, the proposed DS can transform to diverse topological phases by symmetry-lowering perturbations. 

\begin{acknowledgments}
Y.-T.O. was supported from the Global Ph.D. Fellowship Program through the the National Research Foundation of Korea  (NRF) funded by the Ministry of Education (No. NRF-2014H1A2A1018320). 
Y.K. was supported from the NRF grant funded by the Korea government (MSIP; Ministry of Science, ICT \& Future Planning) (No. NRF-2017R1C1B5018169).
The computational resource was provided from the Korea Institute of Science and Technology Information (KISTI).
\end{acknowledgments}

\renewcommand{\thefigure}{S\arabic{figure}}
\setcounter{figure}{0}
\renewcommand{\theequation}{S\arabic{equation}}
\setcounter{equation}{0}

\begin{widetext}
\section{Supplementary Material for ``Dual topological nodal line and nonsymmorphic Dirac semimetal in three dimensions''}
\subsection{First-principles calculations}

Our first-principles calculations were performed based on density functional theory (DFT) as implemented in the \textsc{Quantum Espresso} package \cite{giannozzi09p395502}.  We used the Perdew--Burke--Ernzerhof exchange-correlation functional \cite{Perdew96p3865} and norm--conserving, optimized, designed nonlocal pseudopotentials \cite{Rappe90p1227}.  The spin-orbit coupling was fully considered for the electronic structure calculations using a noncollinear scheme. The electronic wave functions were expanded in terms of a discrete set of plane-waves basis within the energy cutoff of 680 eV.   The 8$\times$8$\times$4  Monkhorst-Pack $\vv k$-points were sampled from the first Brillouin zone (BZ) \cite{Monkhorst76p5188}. The atomic structures were fully relaxed within a force tolerance of 0.005 eV/\AA. 
The lattice constants for relaxed unit cells of BaLaXBY$_5$ family are given in Table \ref{table-relax}.
To highlight the elementary band representation (EBR) of our interest in the DFT calculation of BaLaCuBO$_5$, the Wannier90 package was exploited to construct the tight-binding Hamiltonian by using maximally-localized Wannier function for d$_{xy}$ orbitals of Cu \cite{mostofi2008}.

\begin{table}[]
    \centering
    \begin{tabular}{|l|c|c|c|c|c|c|} 
\hline
         & BaLaCuBO$_5$ & BaLaCuBS$_5$ & BaLaCuBSe$_5$ & BaLaAuBO$_5$ & BaLaAuBS$_5$ & BaLaAuBSe$_5$\\\hline
    a    & 5.4769 \AA & 6.5298 \AA & 6.8915 \AA & 5.7041 \AA & 6.6819 \AA & 7.0001 \AA \\\hline
    c    & 7.4640 \AA & 8.6036 \AA & 8.9960 \AA & 7.7112 \AA & 8.6131 \AA & 9.0063 \AA \\\hline
    \end{tabular}
    \caption{Lattice constants in a and c direction for unit cell of the BaLaXBY$_5$ family.}
    \label{table-relax}
\end{table}

\subsection{Chern number calculations}
\label{sec:topology}
In this section, we introduce two computational methods to calculate the Chern number.
First one is to efficiently find a Weyl point (WP) during the phase transition, which carries a non-zero Chern number. 
Then, we divide the BZ into the cubic-grids and calculate the Berry phase on each surface to track the path of the WP.
The other one is a standard Wilson loop method that we used to determine the Chern number of the time-reversal-invariant plane of the weak-topological insulator (WTI) or the Dirac point (DP).

\figr{fig:z2} illustrates the methods that we employed to calculate the Chern number of WPs during the phase transition. We track the position of WP during the phase transition between trivial and topological insulators, by calculating  
the Berry phase on the surface of the cubic-grid of BZ as in \fig{fig:z2}(a).
Since WP plays the role of the monopole of the Berry connection, a cubic-grid with non-zero Berry phase manifests the WPs of net charge equals to its non-zero Berry phase.
We divide BZs into $400\times400\times400$ cubic-grids to track down the path of WPs as the parameters are changed to complete the phase transition via creation and annihilation of WPs.

On the other hands, the non-abelian Wilson loop calculation provides the technical venue to determine the vanishing Chern number of the DP, as well as $\mathcal{Z}_2$ topological invariant \cite{11yu}.
The Wilson loop has a mathematical structure given by 
\begin{equation}
\mathcal{W}(k_l,\vv k_0) = F(\vv k_0) F(\vv k_1) \cdots F(\vv k_{N-2}) F(\vv k_{N-1}),
\end{equation}
where the overlap matrix $\left[F(\vv k_i)\right]_{nm} \equiv \langle \psi_n(vv k_i) | \psi_m(\vv k_{i+1}) \rangle$ is defined by the inner product of the occupied states at two adjacent momenta $\vv k_i$ and $\vv k_j$ on the closed path $k_l$.
For equal spacing slides of the closed loop $k_l$ with infinitesimal spacing $\Delta \vv k$, the Wilson matrix is mathematically equivalent to the non-abelian Berry phase:
\begin{equation}
\left [\mathcal{W}(k_l, \vv k_0)  \right]_{nm} \equiv P \left[\exp \left( i \oint_{k_l} d\vv k  \cdot \vv A(\vv  k) \right) \right]_{nm},
\end{equation}
where P represents the integral on the closed loop $k_l$ is path-ordered, $\vv k_0$ is the starting point of the closed path integral, and $\left [\vv A(\vv  k)\right]_{nm} \equiv i \langle \psi_n(\vv k) | \partial_{\vv k} \psi_n(\vv k)\rangle$ is the non-abelian Berry connection on the momentum space.

By sweeping the specific momentum plane with the Wilson matrix, one can determine the $Z$ Chern number or $Z_2$ invariant of the plane.
Particularly, in the case of the TR symmetrical plane, the even-number-crossing and odd-number-crossing of $\phi(k_y)$, the phases of eigenvalues of Wilson matrix, indicate the trivial and non-trivial $Z_2$ invariant of the plane \cite{11yu}.
In \fig{fig:z2}(b-1), the Wilson loops on the $k_z = 0$ plane is illustrated.
The Wilson matrix  $\mathcal{W} (k_y)$ is calculated in the closed loops aligned in $k_x$ direction;
\begin{equation}
\mathcal{W}(k_y) = F(k_{x,0},k_y) F(k_{x,1},k_y) \cdots F(k_{x,N-2},k_y) F(k_{x,N-1},k_y),
\end{equation}
where $k_{x,j} = - \pi + 2 \pi j/N$.
\fig{fig:z2}(b-2) shows a clear odd-number-crossing of the eigenphase $\phi(k_y)$, which manifests that the phase is a topological insulator (TI) phase.
We confirm that there exists the WTI phase on the phase diagram generated by the $m_S$ perturbation by implementing the Wilson band calculation on equally separated Wilson loop of $N = 200$.

Additionally, we calculate the Wilson band on the spherical surface enclosing the Weyl nodal line (WNL) to investigate the relation to the DP which is achieved by shrinking the WNL into a point via restoring the inversion symmetry.
The Wilson band sphere enclosing the WNL is parameterized by 
\begin{equation}
\vv k = k_0 \left(\cos \varphi \sin \theta, \sin\varphi \sin \theta, \cos \theta \right),\label{eq:app-wilson_sph} 
\end{equation}
as illustrated in \fig{fig:z2}(c-1).
The Wilson matrix as a function of $\theta$ is calculated via
\begin{equation}
\mathcal{W}(\theta) = F(\theta,\varphi_0) F(\theta,\varphi_1) \cdots F(\theta,\varphi_{N-2}) F(\theta,\varphi_{N-1}),
\label{eq:app-wb-sphere}\end{equation}
where 
$
\left[(\theta,\varphi_{i})\right]_{n,m} = \langle\theta, \varphi_i; n| \theta,\varphi_{i+1}; m\rangle
$
is the overlap matrix of occupied states between two neighboring azimuth angles $\varphi_i  = 2 \pi \,i /N_{\varphi}$ and $\varphi_{i+1}  = 2 \pi \,(i+1) /N_{\varphi}$.
In \fig{fig:z2}(c-2), the phases of eigenvalues of the Wilson matrix, $\phi (\theta)$, are illustrated with adjusting small parameter $k_0$.
The flowing pair of Wilson bands from $0$ to $0$ ($=2$) via $-\pi$ and $\pi$ shows an identical winding structure with the Wilson bands of DP as illustrated in \fig{fig:app-DP}(c).

\begin{figure}[h]
\includegraphics[width=1.\textwidth]{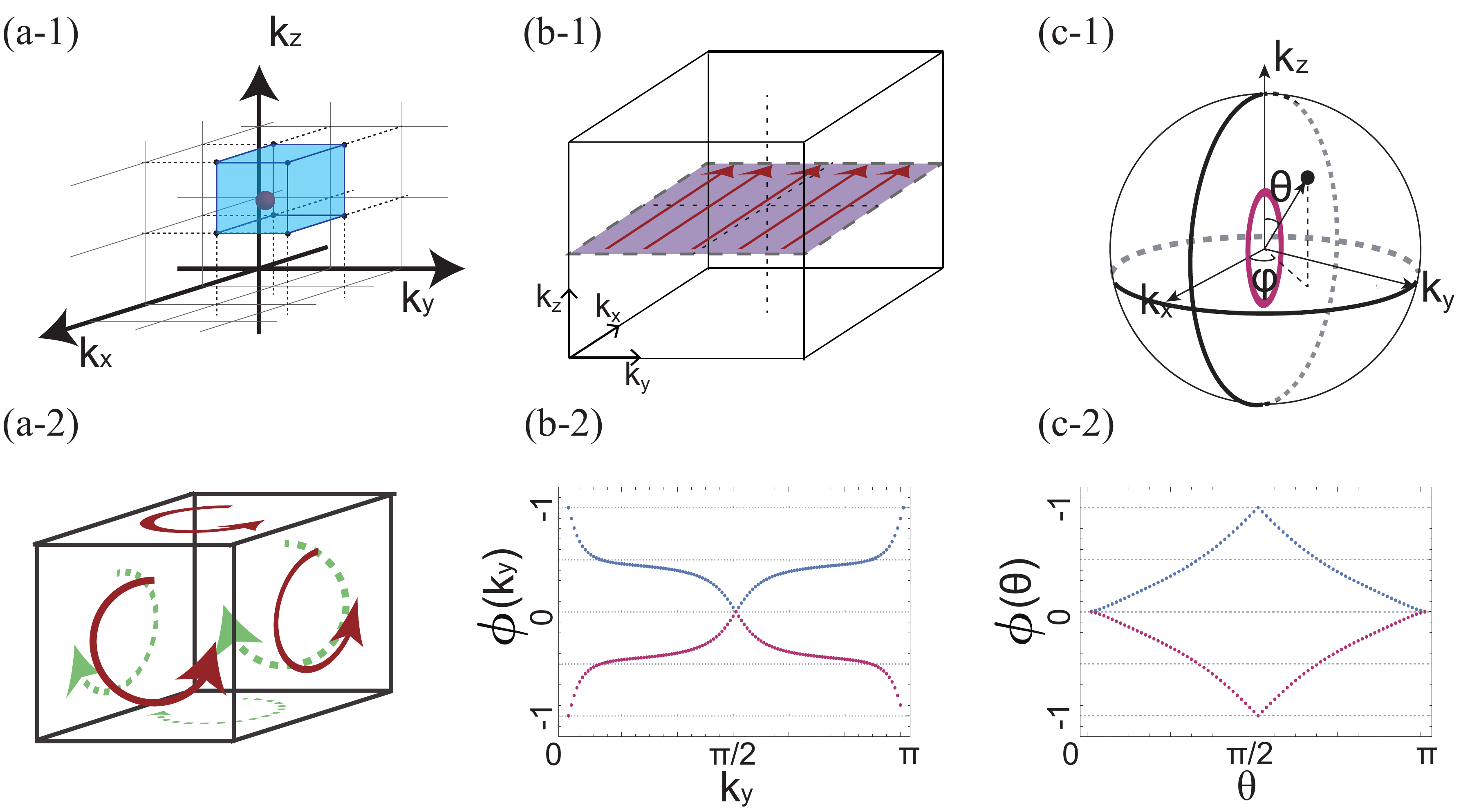}
\caption{
\label{fig:z2} 
(a) Schematic view of the cubic-grid Berry phase calculation for the WP tracking.  (a-1) Blue-shaped cubic denotes the non-zero cubic-grid Berry phase, whereas red dot represents WP inside of it.  (a-2) Berry phase is calculated in the direction as illustrated with red and yellow arrows in the right column.  (b) Illustration of the non-abelian Wilson loop calculation on $k_z = 0$ plane of BZ, and non-trivial Wilson loop result for the WTI phase.  Red arrows on the $k_z = 0$ plane in (b-1) denote the Wilson loops.  Blue and orange dots in (b-2) represent the phases of eigenvalues of Wilson Matrix $\mathcal{W}(k_y)$.  (c) Illustration of the non-abelian Wilson loop calculation on the sphere enclosing the WNL.  The Wilson sphere parameterized by Eq. (\ref{eq:app-wilson_sph}) is illustrated in (c-1) with reasonably small radius $k_0$. Red line in (c-1) represents the WNL located at X (R) point.  Blue and orange dots in (b-2) represent the phases of eigenvalues of Wilson Matrix $\mathcal{W}(\theta)$.
}
\end{figure}

\subsection{Band BaLaXBY$_5$ families}

\begin{figure}[tb]
\includegraphics[width=0.97\textwidth]{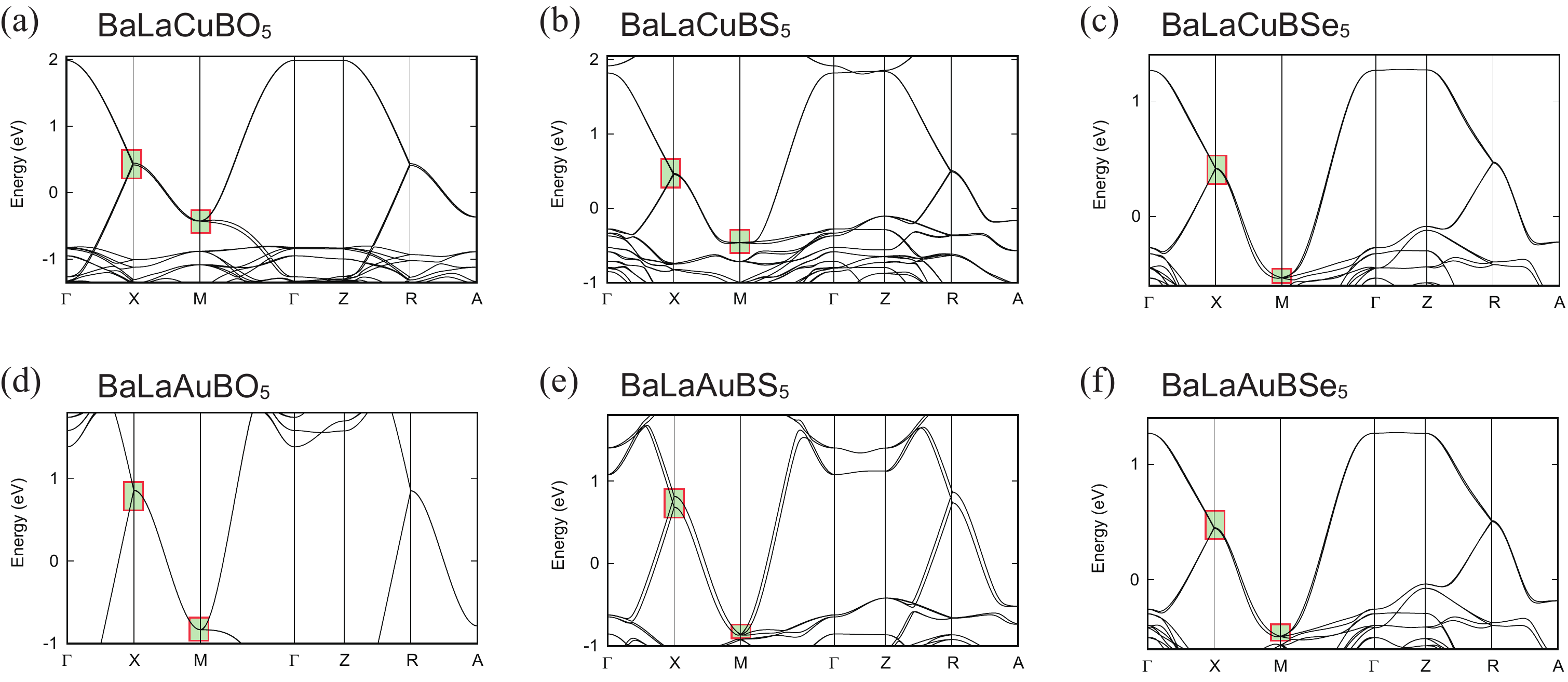}
\caption{\label{fig:app-dft} 
DFT band structures of (a) BaLaCuBO$_5$, (b) BaLaCuBS$_5$, (c) BaLaCuBSe$_5$, (d) BaLaAuBO$_5$, (e) BaLaAuBS$_5$, and (f) BaLaAuBSe$_5$.  In each plot, we emphasize the hourglass-like crossing of doubly-degenerate WNL in the vicinity of $X$ point on the $\Gamma$-$X$ line, and the (DP) at the $M$ point by the red rectangles.
}
\end{figure}

The DFT band structures of BaLa$X$B$Y_5$ family with $X$ = Cu and Au and $Y$ = O, S, and Se, of which the space symmetry belongs to the space group (SG) 100, are shown in \fig{fig:app-dft}.  Every member of this family features the nodal structures with four-band sticking forming and elementary band representations \cite{17bernevig}.  We found that the four bands near the Fermi level mainly comprise the $d$ orbitals of the $X$ atoms.  The Fermi level is well separated from the bands other than the four bands near the Fermi level. The four-band sticking enforces the nodal structure, resulting in a filling-enforced semimetal \cite{watanabe2016} when the filling is $4\mathbb{Z}+2$ as in the cases of these compounds. Near the $X$ point on the $\Gamma$-$X$ line, the hourglass-like crossing appears forming two-fold degenerate node, which constitutes twofold-degenerate WNL forming along the $k_z$ direction. At the same time, the fourfold degenerate DP exists at the $M$ point in all the members of the material family.

\begin{figure}[h]
\includegraphics[width=0.47\textwidth]{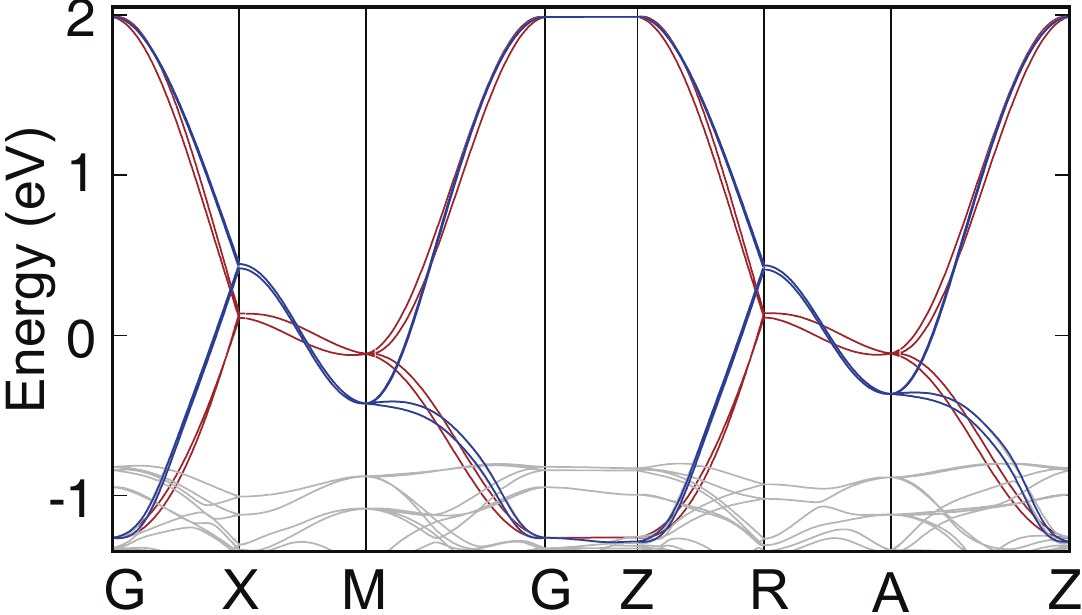}
\caption{\label{fig:tb-app} Band structures for the TB Hamiltonian $\mathcal{H}\left( \vv k \right)$ (red lines) and BaLaCuBO$_5$ from the DFT calculations (blue and gray lines).
DFT result marked with blue color shows analogous structure with TB result. 
}
\end{figure}

\figr{fig:tb-app} shows the comparison between the tight-binding (TB) and first-principles band structure for BaLaCuBO$_5$, where readers can find that the tight-binding (TB) model well reproduces the four DFT bands near the Fermi level.  We provided the TB model in the main text as %
\begin{align}
\mathcal{H}^0({\vv k}) =  &  t_1 \cos \frac{k_x}{2} \cos\frac{k_y}{2} \tau_x +t_2 \left( \cos k_x + \cos k_y  \right)  + t_3 \cos{k_z}  \nn
& + v_0 \cos \frac{k_x}{2} \cos \frac{k_y}{2} \tau_y \sigma_z + v_1 \left( \sin k_x \,  \sigma_x + \sin k_y\,  \sigma_y \right) \tau_z 
+ v_2\left(\sin k_x \, \sigma_y - \sin k_y \, \sigma_x \right) +  v_3 \sin{k_z} \tau_z \sigma_z \nn 
& +  v_4 \left( \sin  \frac{k_x}{2} \, \cos \frac{k_y}{2} \sigma_y 
- \cos  \frac{k_x}{2} \sin  \frac{k_y}{2} \sigma_x \right) \tau_x.
\end{align}
Here, the simple first and second nearest-hoppings designated by $t_1$, $t_2$ are considered in the plane, and along the $k_z$ direction, the nearest-hopping designated by $t_3$ is considered.  The parameter set of $\mathcal{H}^0(\vv k )$ that we find to reproduce the DFT results best is found as $t_0 = 0.124$, $t_1=1.626$, $t_2=0.119$, $t_3=0.0$, $v_0 = 0.0$, $v_1 =0.0$, $v_2 =0.04$, $v_3 = 0.01$, and $v_4 = 0.013$ in the order of eV.  We first matched the energy eigenvalues at the time-reversal-symmetrical momenta (TRIMs), then adjusted the rest of parameters to maximally reduce the mismatches via introducing the interactions beyond the nearest- and next-nearest-neighbor interactions.  

\subsection{Double-glide Dirac node decomposition}
The $M$ point hosts a fourfold degenerate DP, protected by double-glide-mirrors. Here we develop the low-energy effective theory for the Dirac point that respects the symmetries of SG $100$ and the time-reversal (TR) symmetry, represented by %
\begin{equation}
 \mathcal{T} = i \sigma_y \mathcal{K}, ~~
 g_x =  \tau_y \sigma_x , ~~
 C_{4z} =  \tau_z \exp\left[ i \frac{\pi}{4} \sigma_z  \right].
\label{m-generators}
\end{equation}
Note that the phase factors of each representation are arranged, such that they satisfy the commutation relations of SG 100 and possess the corresponding eigenvalues.  The $k \cdot p$ model respecting these symmetries together with the TR symmetry is given by
\begin{equation}
\mathcal{H}^M(\vv k)  = 
k_x   \left( c_1  \tau_z \sigma_x  + c_2 \tau_x \sigma_x   +  c_3  \sigma_y  \right) 
+
k_y \left( c_1  \tau_z \sigma_y  - c_2 \tau_x \sigma_y  - c_3  \sigma_x   \right) 
+  k_z \left( c_4 \tau_z \sigma_z+ c_5 \tau_x \sigma_z \right). \label{eq:kp-sym}
\end{equation}

We introduce five anti-commuting $\Gamma$-matrices and ten combinations of them to express the $SU(4)$ invariant Hilbert space.  Our choice is $\Gamma_{(1,2,3,4,5)} =  (\tau_x \sigma_y, \tau_x \sigma_x,\tau_y ,\tau_x\sigma_z , \tau_z)$, which is followed by their ten combinations of $\Gamma_{ab}\equiv -i\Gamma_{a}\Gamma_{b}~(a\neq b)$.  Note that $\Gamma_{1\sim4}$ are $\mathcal{T}$-odd
\begin{equation}\mathcal{T}^{-1} \Gamma_a\mathcal{T} = -\Gamma_a,\end{equation}
while $\Gamma_{5}$ is $\mathcal{T}$-even
\begin{equation}\mathcal{T}^{-1} \Gamma_5\mathcal{T} = \Gamma_5.\end{equation}
With the $\Gamma$-matrix set, one can rewrite the Hamiltonian of Eq. (\ref{eq:kp-sym}) as following.
\begin{equation}
\mathcal{H}^M (\vv k ) = k_x \left( c_1 \Gamma_{23} + c_2 \Gamma_2 - c_3 \Gamma_{24}\right) 
+  k_y \left( c_1\Gamma_{13} - c_2\Gamma_1 - c_3 \Gamma_{14} \right) 
+  k_z \left( -c_4 \Gamma_{34} + c_5 \Gamma_4 \right).
\end{equation}

The eigenenergies for the Hamiltonian in particular momentum space are given as follows.
\begin{align}
& H(k_x,0,0): \pm \left(c_1^2 + c_2^2 + c_3^2  \right)^{1/2} , \nn
& H(k_x,k_y,0): \pm \left[ \left\{c_2 k_x  \pm \left(c_1^2 + c_3^2\right)^{1/2}k_y\right\}^2 + \left\{c_2 k_y  \pm \left(c_1^2 + c_3^2\right)^{1/2}k_x\right\}^2\right]^{1/2} ,\nn 
& H(k_x,0,k_z): \pm \left\{ c_3^2 k_x^2 + \left(c_2 k_x \pm c_4 k_z \right)^2
+ \left(c_1 k_x \mp c_5 k_z \right)^2  \right\}^{1/2} , \nn
& H(k_x,k_y,k_z):  \nn
& \pm \left[\left(c_1^2 +c_2^2 + c_3^2 \right)\left( k_x^2 + k_y^2 \right)  + \left(c_1^2 + c_5^2\right) k_z^2 
\pm 2 \left\{ 4 c_2^2 \left( c_1^2+c_3^2\right) k_x^2 k_y^2 + \left[\left(c_2c_4 \mp c_1c_5\right)k_x^2 +\left(c_2c_4 \pm c_1c_5\right)k_y^2  \right]k_z^2
\right\}^{1/2}\right]
\end{align}

We find that the symmetry-preserving $k\cdot p$ model is identical with the Taylor-expanded TB result in the vicinity of $M$ point, which is clearly shown by setting the coefficients to $c_1=-v_1$, $c_2 = v_4$, $c_3 = -v_2$, $c_4 = v_3$ , and $c_5 = 0$.
The term $c_5 \ne 0$ is responsible for a higher-order hopping beyond the next-nearest-neighbor interaction, which we excluded in our TB model.
\figr{fig:app-DP}(a) shows the energy-momentum relationship on $k_x-k_y$ plane, obtained from the $k\cdot p$ Hamiltonian. 
Unlike a conventional centrosymmetric Dirac cone, in which all the bands doubly degenerate away from the DP, our double-glide Dirac semimetal (DGDS) DP exhibits bifurcation of the conduction and valence bands away from the DP except on the glide-invariant $k_x = \pi$ and $k_y = \pi$ lines, due to the absence of inversion symmetry. 

\begin{figure}[tb]
\includegraphics[width=0.9\textwidth]{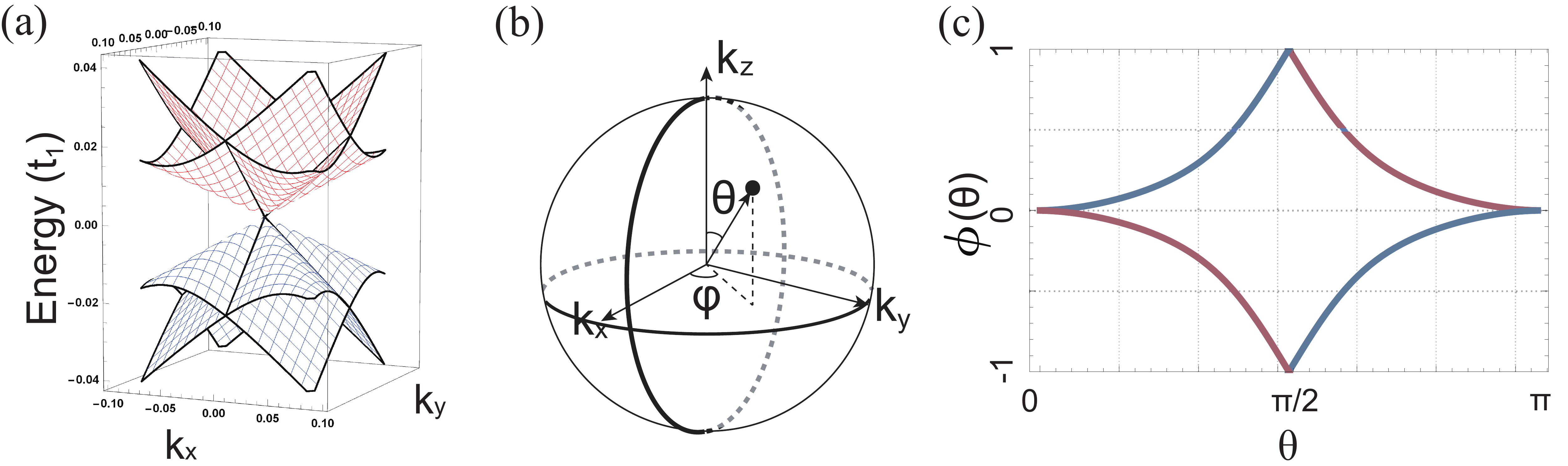}
\caption{\label{fig:app-DP}  
(a) Energy-momentum dispersion around $M$ point. The DP clearly appears at the center.
(b) Wilson-loop vector parameterized by Eq. (\ref{eq:app-wilson_sph}).
(c) Non-abelian Wilson loop calculation of double-glide DP.  The $x$-axis represents zenith angle $\theta$, and the $y$-axis represents the phase of the eigenvalues of the Wilson Hamiltonian. The red (blue) color scheme is used to represent the Wilson loop winding from $0$ to $0$ via $+1$ ($-1$). 
\\
}
\end{figure}


By calculating the Chern number via the Wilson band calculation, we convince ourselves that the DP protected by double-glide-mirrors without inversion symmetry carries zero Chern number, as it is supposed to be since a Dirac cone is a composite of two WPs with opposite Chern numbers $\pm1$. 
The detailed description for the Wilson matrix calculations is provided in Section \ref{sec:topology}.
To evaluate the Chern number of the DP, which is turned out to be zero, 
the Wilson band calculation is implemented on the sphere enclosing the double-glide DP parameterized as in Eq. (\ref{eq:app-wilson_sph}), where the center of the sphere is set to the DP.
The Wilson matrix as a function of $\theta$ is calculated via Eq. (\ref{eq:app-wb-sphere}).
In \fig{fig:app-DP}(c), we plot the phases of eigenvalues of the Wilson matrix, $\phi (\theta)$, with adjusting small parameter $k_0$.
The pair of Wilson bands flows from $0$ to $0$ ($=2$) via $-\pi$ and $\pi$, respectively, which is a typical Chern number calculation for a DP that comprises two WP with opposite Chern number $\pm 1$. 
Our calculation proves that the genuine three-dimensional DP occurs from the $p4g$ multiplayer in SG 100, in which inversion is absence.

\subsection{Symmetry lowering perturbation}
\label{app:pert}

\begin{table}[]
\centering
\begin{tabular}{|c|c|c|c|c|c|c|}
\hline
\multirow{2}{*}{Class}& \multirow{2}{*}{Strain} & \multirow{2}{*}{Perturbation} & \multicolumn{4}{c|}{Band gap open} 
 \\ \cline{4-7}
& & & $\Gamma$&$X$&$Y$&$M$
\\ \hline
\multirow{4}{*}{$A_{2u}$} & \multirow{4}{*}{$\includegraphics[width=0.06\textwidth]{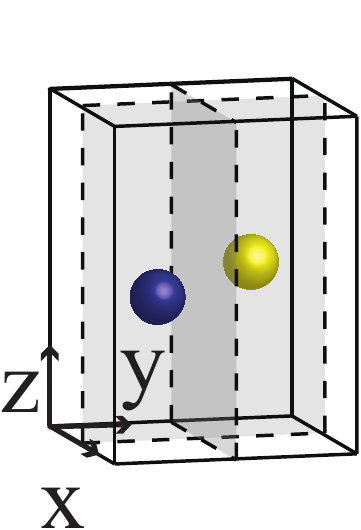} $} & 
\multirow{2}{*}{$\tau_z$}
& \multirow{2}{*}{O} & \multirow{2}{*}{O} & \multirow{2}{*}{O}  & \multirow{2}{*}{O} \\ 
 & & & & & & \\ \cline{3-7}
 & &
\multirow{2}{*}{$\sin\cfrac{k_x}{2}\cos\cfrac{k_y}{2} \tau_x \sigma_x +\cos \cfrac{k_x}{2} \sin \cfrac{k_y}{2} \tau_x \sigma_y$}
 &\multirow{2}{*}{$\times$}&\multirow{2}{*}{O}&\multirow{2}{*}{O}&\multirow{2}{*}{$\times$} \\ 
 & & & & & &\\\hline
\multirow{4}{*}{$B_{1u}$} & \multirow{4}{*}{$\includegraphics[width=0.12\textwidth]{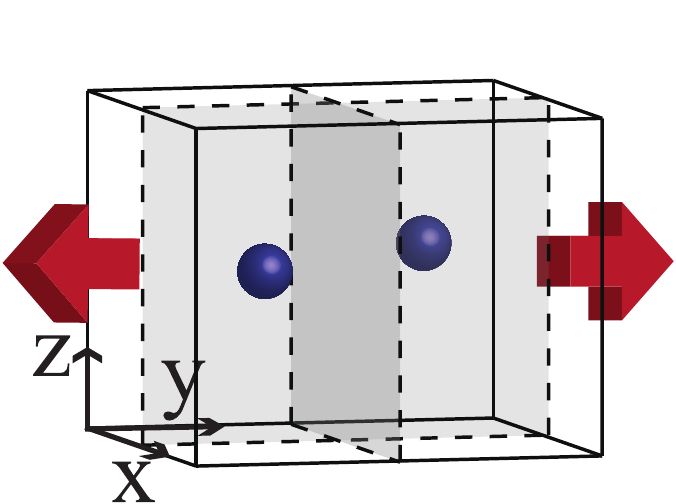}$} &
\multirow{4}{*}{$\sin\cfrac{k_x}{2}\cos\cfrac{k_y}{2} \tau_x \sigma_y  -\cos \cfrac{k_x}{2} \sin \cfrac{k_y}{2} \tau_x \sigma_x$}
&\multirow{4}{*}{$\times$}& \multirow{4}{*}{O} & \multirow{4}{*}{O}&\multirow{4}{*}{ $\times$ } \\ 
 & & & & &  &\\ 
 & & & & &  &\\
 & & & & &  &\\ \hline
\multirow{4}{*}{$B_{2g}$} & \multirow{4}{*}{$\includegraphics[width=0.12\textwidth]{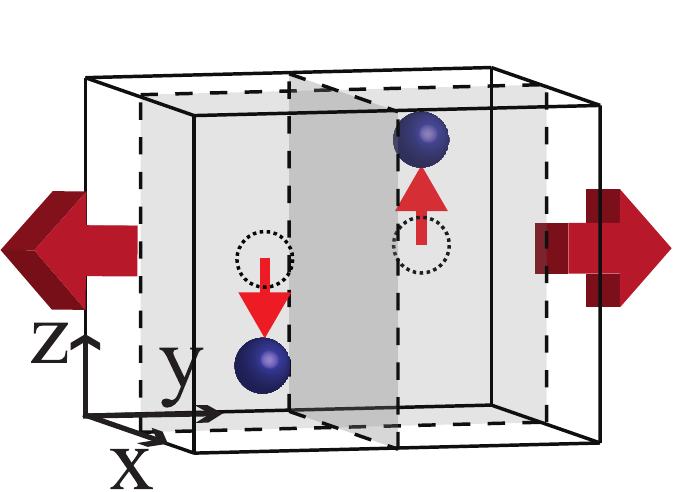}$} & 
\multirow{4}{*}{$\sin\cfrac{k_x}{2}\sin\cfrac{k_y}{2} \tau_x$}
&\multirow{4}{*}{ $\times$ } & \multirow{4}{*}{$\times$} & \multirow{4}{*}{$\times$} & \multirow{4}{*}{O} \\ 
 & & & & & & \\
 & & & & & & \\
 & & & & & & \\\hline
\multirow{4}{*}{$B_{2u}$} & \multirow{4}{*}{$\includegraphics[width=0.12\textwidth]{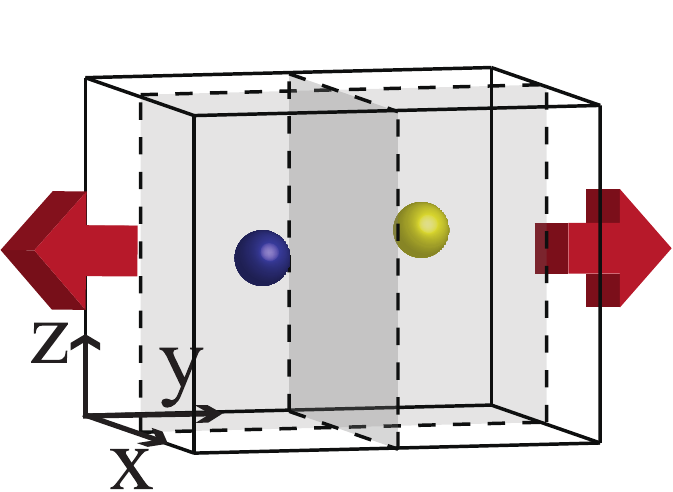}$} &
\multirow{4}{*}{$\sin\cfrac{k_x}{2}\cos\cfrac{k_y}{2} \tau_x \sigma_x  -\cos \cfrac{k_x}{2} \sin \cfrac{k_y}{2} \tau_x \sigma_y$}
&\multirow{4}{*}{$0$} & \multirow{4}{*}{$\tau_x \sigma_x$} & \multirow{4}{*}{$- \tau_x \sigma_y $} & \multirow{4}{*}{$\tau_y \sigma_z$} \\ 
 & & & & & &\\
 & & & & & &\\
 & & & & & &\\\hline
\multirow{4}{*}{$E_{g}$} & \multirow{4}{*}{$\includegraphics[width=0.12\textwidth]{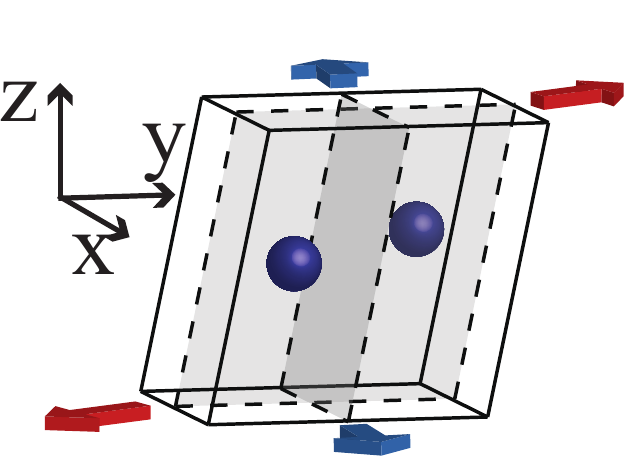}$} &
\multirow{2}{*}{$\sin\cfrac{k_x}{2}\cos\cfrac{k_y}{2} \tau_y$}
& \multirow{2}{*}{$\times$} & \multirow{2}{*}{O} & \multirow{2}{*}{$\times$} & \multirow{2}{*}{$\times$} \\ 
 & & & & & &\\\cline{3-7}
 & &
\multirow{2}{*}{$\cos\cfrac{k_x}{2}\sin\cfrac{k_y}{2} \tau_y$}
 & \multirow{2}{*}{$\times$} & \multirow{2}{*}{$\times$} & \multirow{2}{*}{O} & \multirow{2}{*}{$\times$}\\
 & & & & & &\\\hline
 & & 
 $\cos\cfrac{k_x}{2} \cos\cfrac{k_y}{2} \tau_y \sigma_x$ 
 & O & $\times$ & $\times$ & $\times$ \\\cline{3-7}
  & \multirow{4}{*}{$\includegraphics[width=0.12\textwidth]{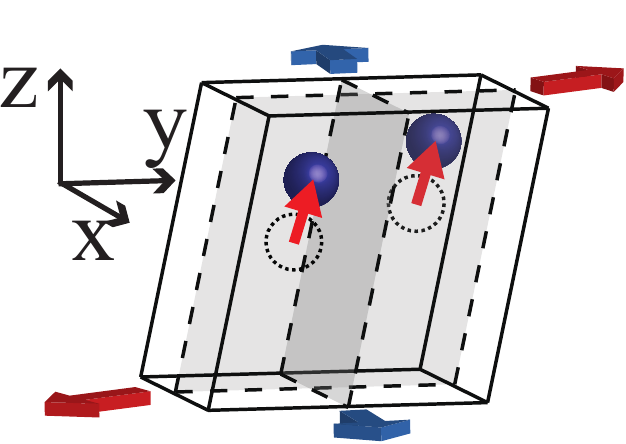} $} &
 $\sin\cfrac{k_x}{2} \sin\cfrac{k_y}{2} \tau_y \sigma_y$
 & $\times$ & $\times$ & $\times$ & O \\\cline{3-7}
 \multirow{2}{*}{$E_{u}$} & &
 $\sin\cfrac{k_x}{2} \cos\cfrac{k_y}{2} \tau_x \sigma_z$
 & $\times$ & O & $\times$ & $\times$ \\\cline{3-7}
 & &
 $\cos\cfrac{k_x}{2} \cos\cfrac{k_y}{2} \tau_y \sigma_y$ 
 & O & $\times$ & $\times$ & $\times$ \\\cline{3-7}
  & &
 $\sin\cfrac{k_x}{2} \sin\cfrac{k_y}{2} \tau_y \sigma_x$
 & $\times$ & $\times$ & $\times$ & O \\\cline{3-7}
   & &
 $\cos\cfrac{k_x}{2} \sin \cfrac{k_y}{2} \tau_x \sigma_z$
 & $\times$ & O & $\times$ & $\times$ \\\hline
\end{tabular}
\caption{
Classification and representations of symmetry-lowering perturbations of SG $100$. 
The unit cell is illustrated by a box. 
Different color schemes are used for the atoms to represent different on-site energy.
}
\label{tb-4b-pert}
\end{table}

In this Section, we provide detailed information about symmetry lowering perturbations of SG 100.
In Table \ref{tb-4b-pert}, we classify the possible symmetry lowering perturbations in the minimal model of SG 100.
Note that we consider only the perturbations that open the gap of at least one TRIM.
In the first column, we classify the perturbations by the symmetry representations of the point group D$_{4h}$.
Note that the subscript $u$ ($g$) is introduced to designate the inversion symmetric (asymmetric) perturbation. 
We provide the atomistic illustrations in the second column for the perturbations that cause the corresponding symmetry-lowering perturbation, which can be considered as a combination of uni-axial strain, buckling, and staggered potential.
We also provide the representation of the perturbation in the third column. In the last four columns, we inform whether the perturbation opens an energy band gap at the corresponding TRIM, $\Gamma$, $X$, $Y$, and $M$.
In the case where multiple perturbations are allowed in a class, they are separate with horizontal lines.

The perturbation of the first class $A_{2u}$ breaks double-glide-mirrors $g_x$ and $g_y$ but preserves the $C_{4z}$ symmetry. 
The different on-site energies of the two sublattices can be achieved by substituting one of the two sublattices.
On the other hands, $B_{1u}$ class breaks $C_{4z}$, while preserving double-glide-mirrors $g_x$ and $g_y$, which can be achieved by applying a uniaxial strain.
Both $B_{2g}$ and $B_{2u}$ classes break $C_{4z}$ and double-glide-mirrors $g_x$ and $g_y$. The only difference between $B_{2g}$ and $B_{2u}$ is inversion symmetry;  $B_{2g}$ additionally breaks the inversion symmetry, while $B_{2u}$ preserves it.
The combination of uniaxial strain and inversion-preserving buckling or inversion-breaking substitution can generate $B_{2g}$ and $B_{2u}$, respectively.
$E_g$ and $E_u$ classes preserve one of the double-glide-mirrors, $g_x$ or $g_y$ and break the other glide as well as $C_{4z}$ symmetry.
$E_g$ can be achieved applying shear stress in the $x$-$z$ or $y$-$z$ plane, while additional glide-preserving buckling generates $E_u$ by breaking inversion symmetry.

For the phase diagram, $A_{2u}$, $B_{2g}$ and $E_g$ classes of the perturbations are considered.
To study the effect of breaking glide-mirrors of the DGDS, we consider the class $A_{2u}$, which breaks double-glide-mirrors selectively, and $m_{\protect\scalebox{.55}{$A_{\protect\scalebox{1.}{$2u$}}$}} \tau_x$ is chosen since it affects the entire BZs.
Also, $B_{2g}$ and $E_{g}$ classes are considered with the expectation that they drive DGDS to the (weak) TI phase.
The perturbed Hamiltonian for these perturbations can be written as
\begin{align}
H^1(\vv k) = m_{\protect\scalebox{.55}{$A_{\protect\scalebox{1.}{$2u$}}$}} \tau_x + m_{\protect\scalebox{.55}{$E_{\protect\scalebox{1.}{$g$}}$}}^{(1)} \sin\frac{k_x}{2} \cos\frac{k_y}{2} \tau_y +  m_{\protect\scalebox{.55}{$E_{\protect\scalebox{1.}{$g$}}$}}^{(2)} \cos \frac{k_x}{2} \sin\frac{k_y}{2} \tau_y + m_{\protect\scalebox{.55}{$B_{\protect\scalebox{1.}{$2g$}}$}} \sin\frac{k_x}{2}\sin\frac{k_y}{2}.
\end{align}
By adding the perturbation $H^1(\vv k)$ to the pristine Hamiltonian $H^0(\vv k) = H^0_+(\vv k) + H^0_-(\vv k)$, we study the phase diagram achievable via applying the corresponding symmetry-lowering perturbations. Note that we divide the $H^0_+(\vv k)$ and $H^0_-(\vv k)$ are the centrosymmetric and non-centrosymmetric parts of the pristine Hamiltonian are given respectively by
\begin{align}
\mathcal{H}^{0}_+(\vv k)  =  t_1 \cos\frac{k_x}{2} \cos\frac{k_y}{2} \tau_x + v_1 \left(\sin k_x \tau_z \sigma_x + \sin k_y \tau_z \sigma_y \right)  
+ v_3 \sin{k_z} \tau_z \sigma_z, \nonumber
\end{align}
and
\begin{align}
\mathcal{H}^{0}_-(\vv k)  =  v_- \Bigg[ v_0 \cos\frac{k_x}{2} \cos\frac{k_y}{2} \tau_y \sigma_z +  v_2\left(\sin k_x \, \sigma_y - \sin k_y \, \sigma_x \right)   
 + v_4 \left( \sin  \frac{k_x}{2} \, \cos \frac{k_y}{2}  \sigma_y  - \cos  \frac{k_x}{2} \sin  \frac{k_y}{2}  \sigma_x \right)\tau_x  \Bigg].
\label{eq:unperturbed-h}
\end{align}

Similar to a regular nonsymmorphic Dirac semimetal (DS), which defines a critical point between normal insulator (NI) and  topological insulator (TI) phases \cite{murakami-07-njp,12young,young15prl}, the DGDS occurs at a phase boundary between NI and TI phases tuned by symmetry-lowering perturbations. We demonstrated this by explicitly calculating $\mathcal{Z}_2$ topological invariant when a band gap is opened by symmetry-lowering perturbations.
To efficiently calculate the $\mathcal{Z}_2$ topological invariant, we start from the centrosymmetric limit and push it into the noncentrosymmetric limits until a band gap closes and reopens, which signals a topological phase transition. 
In the centrosymmetric limit, one can easily determine the TI phase by calculating the Fu-Kane $\mathbb{Z}_2$ index $(\nu_1, \nu_2, \nu_3 ; \nu)$, which can be calculated as
\begin{equation} \nu_n=\prod_i \xi_i , \end{equation} 
where product $i$ runs over TRIMs at high-symmetrical plane $j$ of BZ, and $\xi_i$ corresponds to the inversion-symmetry eigenvalues \cite{07fu-kane-mele}.
The absence of band gap closing guarantees  the same topological insulator state that we find from the centrosymmetric limit in the noncentrosymmetric region. 
The inversion symmetric limit can be obtained by setting $v_-= 0$ and $ m_{\protect\scalebox{.55}{$A_{\protect\scalebox{1.}{$2u$}}$}}= 0$, where every band is doubly degenerate due to the Kramers theorem.
One can open the energy gap at TRIMs by setting $t_1$, $m_{\protect\scalebox{.55}{$E_{\protect\scalebox{1.}{$g$}}$}}^{(1)}$, $m_{\protect\scalebox{.55}{$E_{\protect\scalebox{1.}{$g$}}$}}^{(2)}$ and 
$m_{\protect\scalebox{.55}{$B_{\protect\scalebox{1.}{$2g$}}$}}$ nonzero.
Without considering any accidental band crossing between the occupied and conduction bands off high-symmetry momenta, the Kane-Mele $\mathbb{Z}_2$ invariant becomes $(0,0,\zeta;0)$, where $\zeta$ is given by
\begin{equation}
\zeta = -{\rm Sign} \left(t_1 \right) \times {\rm Sign} \left( m_{\protect\scalebox{.55}{$E_{\protect\scalebox{1.}{$g$}}$}}^{(1)} \right) \times {\rm Sign} \left( m_{\protect\scalebox{.55}{$E_{\protect\scalebox{1.}{$g$}}$}}^{(2)} \right) \times  {\rm Sign} \left( m_{\protect\scalebox{.55}{$B_{\protect\scalebox{1.}{$2g$}}$}}\right).
\label{eq:app-KM-index}\end{equation}
Here, the representations for inversion symmetry at TRIMs $\Gamma$, $X$, $Y$ and $M$ are given by
\begin{align}
I_\Gamma = \tau_x , ~~~ I_X = -\tau_y, ~~~ I_Y = \tau_y, ~~~ I_M = \tau_x.
\label{eq:z2-m}\end{align}
We confirm that Eq. (\ref{eq:app-KM-index}) is consistent with the $\mathbb{Z}_2$ invariant calculated by the non-abelian Wilson loop as illustrated in \fig{fig:z2}(b).
Since only the signs of the parameters $m_{\protect\scalebox{.55}{$E_{\protect\scalebox{1.}{$g$}}$}}^{(1)}$, $m_{\protect\scalebox{.55}{$E_{\protect\scalebox{1.}{$g$}}$}}^{(2)}$ and $ m_{\protect\scalebox{.55}{$B_{\protect\scalebox{1.}{$2g$}}$}}$ matters to the $\mathbb{Z}_2$ invariant, we set $m_{\protect\scalebox{.55}{$E_{\protect\scalebox{1.}{$g$}}$}}^{(1)}=m_{\protect\scalebox{.55}{$E_{\protect\scalebox{1.}{$g$}}$}}^{(2)}=m_{\protect\scalebox{.55}{$E_{\protect\scalebox{1.}{$g$}}$}}$ as a representative system for simplicity.
Then, the perturbation in the Hamiltonian becomes 
\begin{align}
\mathcal{H}^1 (\vv k) = m_{\scalebox{.55}{$E_{\scalebox{1.}{$g$}}$}} \sin\left( \frac{k_x+k_y}{2}\right) \tau_y  + m_{\scalebox{.55}{$B_{\scalebox{1.}{$2g$}}$}} \sin\frac{k_x}{2}\sin\frac{k_y}{2} \tau_x  
+ m_{\scalebox{.55}{$A_{\scalebox{1.}{$2u$}}$}} \tau_z, \label{eq:h-pert-part}
\end{align}
Also, we further simplify the Hamiltonian by setting $\mathcal{H}^1 (\vv k) = m_{\scalebox{.55}{$E_{\scalebox{1.}{$g$}}$}} = m_{\scalebox{.55}{$B_{\scalebox{1.}{$2g$}}$}} = m_S$, of which the results are presented in the main text.
In the following Section, the detail description of the phase diagrams in terms of parameters $v_{-}$,  $m_s$, $m_{\protect\scalebox{.55}{$A_{\protect\scalebox{1.}{$2u$}}$}}$  is  provided.

\subsection{\texorpdfstring{$v_{-}$,  $m_s$, $m_{\protect\scalebox{.55}{$A_{\protect\scalebox{1.}{$2u$}}$}}$}{TEXT} phase diagram}
\label{app:phase_pert}

\begin{figure}[tb]
\includegraphics[width=0.97\textwidth]{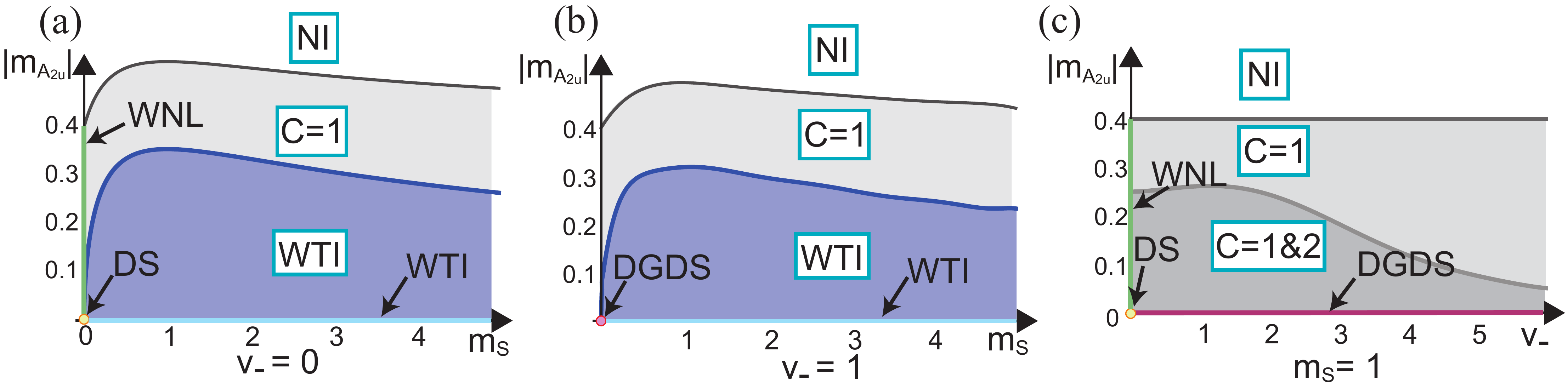}
\caption{\label{fig:app-v-ms-mA} 
The phase diagram of the control parameter $v_{-}$, $m_s$ and $m_{\protect\scalebox{.55}{$A_{\protect\scalebox{1.}{$2u$}}$}}$ with the constraint parameter set $\{t_1, t_2, t_3, v_0, v_1, v_2, v_3, v_4\}  = \{0.35, 0.01,0.02,0.05,0.5,0.1,0.4,0.45\}$.
(a-b) $m_s$-$m_{\protect\scalebox{.55}{$A_{\protect\scalebox{1.}{$2u$}}$}}$ phase diagrams for $v_- = 0$ and $v_-=1$, and (c) $v_-$-$m_{\protect\scalebox{.55}{$A_{\protect\scalebox{1.}{$2u$}}$}}$ phase diagram.
(a) On the $m_{\protect\scalebox{.55}{$A_{\protect\scalebox{1.}{$2u$}}$}}$-axis, the WNL phase is marked with bold-green line, which sprouts from the DS phase colored with orange.
White, gray, and blue areas represent NI phase, $\mathcal{C} = 1$ Weyl semimetal (WS) phase, and WTI phase, respectively. 
(b) Overall structure resembles with (a), but slight differences exist such that DS phase transformed into the DGDS phase, and the WNL phase has vanished.
(c) Identical WNL phase and DS phase of those in (a) are marked with the green line and the orange dot, respectively.
WS phases of $\mathcal{C}=1$ and $\mathcal{C}=1\,\&\, 2$ are separated by gray level.
Red lines on $v_-$-axis represents the DGDS phase.
}
\end{figure}

Here, we provide detailed results of the perturbed Hamiltonian $\mathcal{H}'(\vv k) = H^0(\vv k)+H^1(\vv k)$ in the DGDS and establish phase diagrams generically accessible from the DGDS, where $ H^0(\vv k)$ and $H^1(\vv k)$ are defined in Eqs. (\ref{eq:unperturbed-h}) and  (\ref{eq:h-pert-part}).
In \fig{fig:app-v-ms-mA}, 2D version of the phase diagram for parameters $v_{-}$,  $m_s$ and $m_{\protect\scalebox{.55}{$A_{\protect\scalebox{1.}{$2u$}}$}}$ are given.
As illustrated in \fig{fig:app-v-ms-mA}(a) and \ref{fig:app-v-ms-mA}(b), the $m_s- m_{\protect\scalebox{.55}{$A_{\protect\scalebox{1.}{$2u$}}$}}$ phase diagrams for $v_{-} = 0$ and $v_{-} = 1$ share a similar sketch, with few differences in detail.
In common, the white, gray, and blue colored areas correspond to normal insulator (NI), $\mathcal{C}=1$ WS, and WTI phases, respectively.
In contrast, nonsymmorphic DS (orange dot) and WNL (green line) phase are found in $v_{-} = 0$ limit, while DGDS phase replaces the nonsymmorphic DS and the WNL phase vanishes when $v_{-} = 1$. 
The remaining uncolored white area represents the NI phase.
All the phases, NI, WS, WNL, nonsymmorphic DS, and DGDS phases, are recovered in the 
$v_{-}- m_{\protect\scalebox{.55}{$A_{\protect\scalebox{1.}{$2u$}}$}}$ 
phase diagram in \fig{fig:app-v-ms-mA}(c), only WS phase is distinguished by a different Chern number.

We analyze the symmetry group of which the perturbed system satisfied.
We start by the unperturbed system in the inversion symmetric limit, where $\tau_x$ is chosen for the representation of (restored) inversion.
The corresponding symmetry group belongs to SG 125, where the nonsymmorphic DS phase resides.
The perturbation $m_{\protect\scalebox{.55}{$A_{\protect\scalebox{1.}{$2u$}}$}}$ breaks the inversion and glide-mirrors in the way that the two-fold rotations $C_{2x,2y}$ survive, and brings the system to SG 89 where WNL phase can exist.
On the other hands, increasing $m_S$ from SG 129 brings the system to SG 13, of which the generators are the inversion $I$ and the glide $g_z$.
In SG 13, the WTI phase can be obtained by band inversion.

%
\begin{figure}[h]
\includegraphics[width=0.9\textwidth]{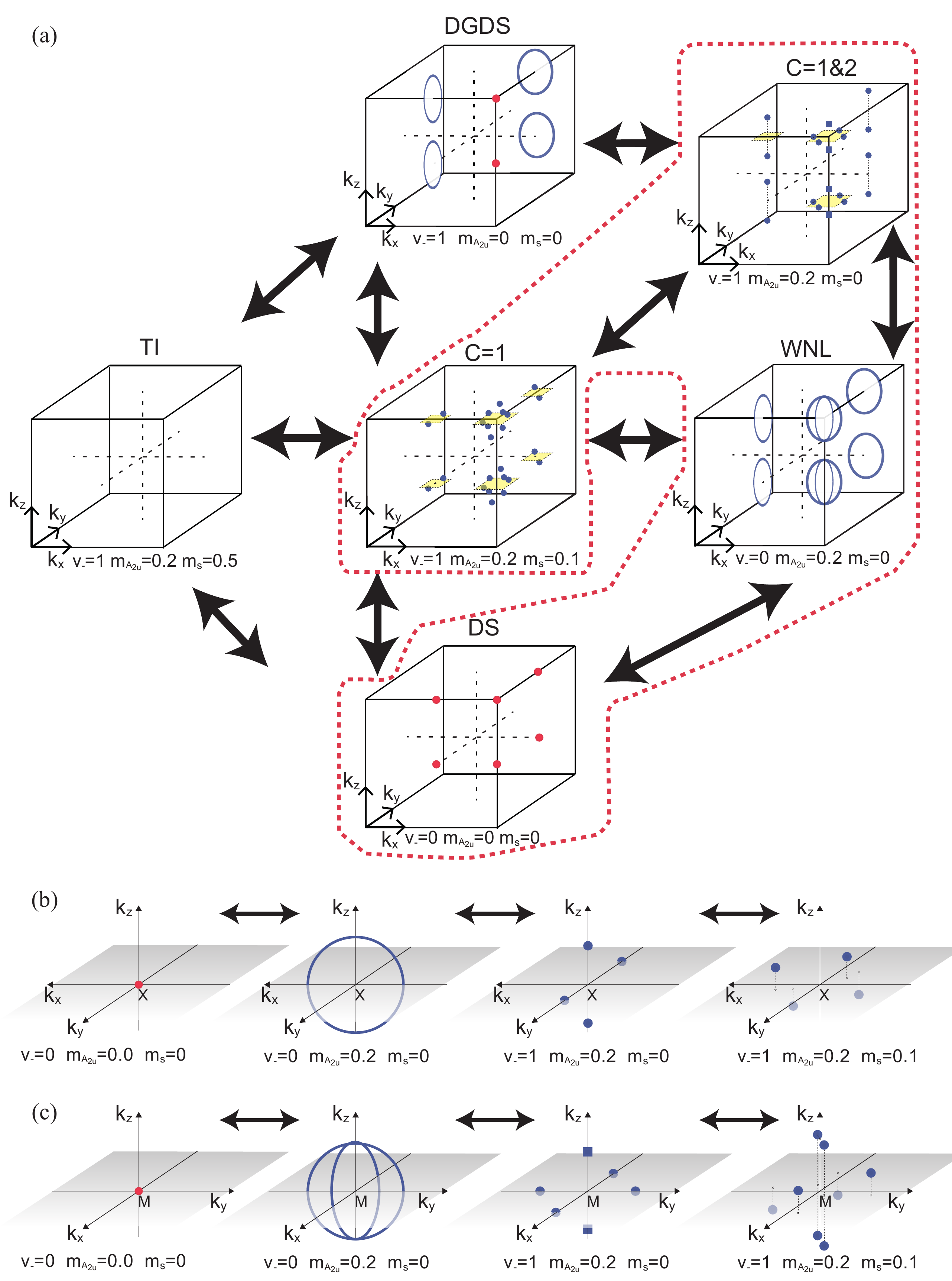}
\caption{
\label{fig:app-nodal} 
Three-dimensional (3D) sketches of the nodal structures in the BZ of the phases DS, DGDS, and WNL phases of \fig{fig:app-v-ms-mA}.
Round red dots, blue lines, and round and square blue dots represent the DP, WNL, and $\mathcal{C}=1$ and $2$ WPs, respectively.
(a) Neighboring relations of the phases in the diagram.
The repeated nodal structures due to the $2 \pi$ periodic structure of BZ are omitted.
For  $\mathcal{C}=1$ and $2$ WS phases, dashed rhombuses are added to help the 3D illustration of the WPs.
(b) Detail illustration of the nodal structure near $X$ points on the phase transition path in the red box of (a).
(c) Detail illustration of the nodal structure near $M$ points on the phase transition path in the red box of (a).
}
\end{figure}

In \fig{fig:app-nodal}, the sketches of the nodal structures of each phase in \fig{fig:app-v-ms-mA} deforming each other are given.
\fig{fig:app-nodal}(a) shows the entire nodal structures of the whole BZ, while detailed local illustrations in the vicinity of $X$ ($R$) and $M$ ($R$) points are given in \figs{fig:app-nodal}(c) and \ref{fig:app-nodal}(d).
The DPs can be achieved by compressing the WNLs, which corresponds to turning off $m_{\protect\scalebox{.55}{$A_{\protect\scalebox{1.}{$2u$}}$}}$ from the WNL phase as illustrated in (b) and (c).
The DP at $X$ ($R$) point is the compression of single WNL while DP at $M$ ($A$) is the compression of two WNL crossing at $k_z$-axis.
Note that WNL at $X$ ($R$) point the DGDS phase deforms to the DP of DS phase in the same way of (b).

On the other hands, increasing $v_-$ from the WNL phase shrinks the WNL into WPs of which the positions are located on the trace of the WNL.
WNL at $X$ ($R$) is deformed into two $\mathcal{C}=-1$ ($\mathcal{C}=1$) on the crossing points of WNL and $k_z$-axis and two $\mathcal{C}=1$ ($\mathcal{C}=-1$) WPs on the $k_z=0$ ($k_z = \pi$) plane, while WNL at $M$ ($A$) point is deformed into two $\mathcal{C}=2$ ($\mathcal{C}=-2$) WPs on the crossing points of the WNL and $k_z$-axis and four $\mathcal{C}=-1$ ($\mathcal{C}=1$) WPs on the crossing points of the WNL and $k_z= 0$ ($k_z = \pi$) plane.
Increasing $m_s$ form the WS phase separate each $\mathcal{C}=2$ WP into two $\mathcal{C}=1$ WPs, respectively, then yields the complex movements of WPs as illustrated in the main text and ends up with the WTI phase.
Indeed, in general, the phase transition occurs along a path other than those of in \fig{fig:app-nodal}, but the basic formula of the deformation of the nodal structures follows the sketches in \fig{fig:app-nodal}.

\end{widetext}
\bibliography{refs}

\end{document}